\documentclass[12pt,a4paper]{article}
\pdfoutput=1

\usepackage{geometry}
\usepackage{amsmath}
\usepackage{amssymb}
\usepackage[dvips]{graphicx}
\usepackage{cite}
\usepackage{pstricks}
\usepackage{bm}
\usepackage{pbox}
\usepackage{placeins}
\usepackage{graphicx}
\usepackage{caption}
\usepackage{subcaption}
\usepackage[T1]{fontenc}
\usepackage{footnote}
\usepackage{pdfpages}
\usepackage{hhline}
\usepackage{multirow}
\usepackage{multicol}
\usepackage{enumitem}
\usepackage[toc,page]{appendix}
\allowdisplaybreaks

\definecolor{MyDarkBlue}{rgb}{0.1, 0.1, 0.8} 
\definecolor{SBlue}{rgb}{0.2, 0.4, 0.7} 
\definecolor{MyLightBlue}{rgb}{0.22,0.51,0.9}
\definecolor{MyGreen}{rgb}{0.0, 0.5, 0.0}
\definecolor{BrickRed}{rgb}{0.8, 0.25, 0.33}
\usepackage[colorlinks=true,linkcolor=blue,citecolor=MyDarkBlue,
urlcolor=MyLightBlue,bookmarksnumbered=true,bookmarksopen]{hyperref}
\hypersetup{colorlinks, citecolor=BrickRed,linkcolor=MyDarkBlue, urlcolor=MyLightBlue}

\begin{document}
\vspace*{-0.2in}
\begin{flushright}
\color{SBlue}{\bf OSU-HEP-19-03}
\end{flushright}
\vspace{0.5cm}
\begin{center}
{\Large\bf 
Minimal Dirac Neutrino Mass Models from  $\bf{U(1)_R}$ Gauge \\ \vspace{0.1in} Symmetry  and  Left-Right Asymmetry at Colliders}
\end{center}

\vspace{0.5cm}
\renewcommand{\thefootnote}{\fnsymbol{footnote}}
\begin{center}
{\large
{}~\textbf{Sudip Jana}\footnote{ E-mail: \textcolor{MyLightBlue}{sudip.jana@okstate.edu}},
{}~\textbf{Vishnu P.K.}\footnote{ E-mail: \textcolor{MyLightBlue}{vipadma@okstate.edu}}
{}~\textbf{and Shaikh Saad}\footnote{ E-mail: \textcolor{MyLightBlue}{shaikh.saad@okstate.edu}}
}
\vspace{0.5cm}

{\em Department of Physics, Oklahoma State University, Stillwater, OK 74078, USA
} 
\end{center}

\renewcommand{\thefootnote}{\arabic{footnote}}
\setcounter{footnote}{0}
\thispagestyle{empty}

\begin{abstract}
In this work, we propose minimal realizations for  generating Dirac neutrino masses in the context of a right-handed  abelian gauge extension of the Standard Model. Utilizing only $U(1)_R$ symmetry, we address and analyze the possibilities of Dirac neutrino mass generation via (a) \textit{tree-level seesaw} and (b) \textit{radiative correction at the  one-loop level}. One of the presented radiative models implements the attractive   \textit{scotogenic} model that links neutrino mass with Dark Matter (DM), where the stability of the DM is guaranteed from a residual discrete symmetry emerging from $U(1)_R$.  Since only the right-handed fermions carry non-zero charges under the $U(1)_R$, this framework leads to sizable and distinctive Left-Right  asymmetry as well as Forward-Backward  asymmetry discriminating from $U(1)_{B-L}$ models and can be tested at the colliders. We analyze the current experimental bounds and present the discovery reach limits for the new  heavy gauge boson $Z^{\prime}$ at the LHC and ILC. Furthermore, we also study  the associated charged lepton flavor violating processes, dark matter phenomenology and cosmological constraints of these models. 
\end{abstract}
\newpage
\setcounter{footnote}{0}

{
  \hypersetup{linkcolor=black}
  \tableofcontents
}
\newpage

\section{Introduction}\label{SEC-01}
Neutrino oscillation data \cite{data} indicates that at-least two neutrinos have tiny masses. The origin of the neutrino mass is one of the unsolved mysteries in Particle Physics. The minimal way to obtain the non-zero  neutrino masses is to introduce three right-handed neutrinos that are  singlets under the Standard Model (SM). Consequently, Dirac neutrino mass term  at the tree-level is allowed and has the form: $\mathcal{L}_Y\supset y_{\nu}\overline{L}_L\widetilde{H}\nu_R$.  However, this leads to unnaturally small  Yukawa couplings  for neutrinos ($y_{\nu} \leq10^{-11}$). There have been many proposals  to naturally induce neutrino mass  mostly by using the seesaw mechanism \cite{typeI, typeII, typeIII, inverse, Bertuzzo:2018ftf} or via radiative mechanism \cite{loop}. Most of the models of neutrino mass generation assume that the neutrinos are Majorana \footnote{For a recent review on models based on Majorana neutrinos see Ref.  \cite{Cai:2017jrq}. For Majorana neutrino mass models within the context of simple grand unified theories see Ref.  \cite{Saad:2019vjo}.}  type in nature. Whether neutrinos are Dirac or Majorana type particles is still an open question. This issue can be resolved by neutrinoless double beta decay experiments  \cite{neutrinoless}.  However, up-to-now there is no concluding evidence from these experiments.

Recently, there has been a growing interest in models where neutrinos are assumed to be Dirac particles. Many of these models use ad hoc discrete symmetries \cite{Ma:2014qra, Ma:2015mjd, Bonilla:2016zef, Chulia:2016giq, Bonilla:2016diq, Ma:2016mwh, Wang:2016lve, Chulia:2016ngi, Borah:2017dmk, Yao:2018ekp, Reig:2018mdk}   to forbid the aforementioned unnaturally small  tree-level Yukawa term  as well as  Majorana mass terms. However, it is more appealing to forbid all these unwanted terms utilizing simple gauge extension of the SM instead of imposing  discrete or continuous global symmetries. This choice is motivated by the fact that contrary to gauge symmetries, global symmetries are known not to be respected by the gravitational interactions \cite{Giddings:1987cg, Giddings:1989bq, Giddings:1988wv, Abbott:1989jw, Coleman:1989zu}. 

In this work, we extend the SM with $U(1)_R$ gauge symmetry, under which only the SM  right-handed fermions are charged and the left-handed fermions transform trivially.
This   realization is very simple in nature and has several compelling features to be discussed in great details. Introducing only the three right-handed neutrinos all the gauge anomalies can be canceled  and  $U(1)_{R}$ symmetry can be utilized to  forbid all the unwanted terms to build desired models of Dirac neutrino mass. Within this framework, by employing the  $U(1)_R$ symmetry we construct a tree-level Dirac seesaw model \cite{Roy:1983be}  and two models where neutrino mass appears at the one-loop level. One of these loop models presented in this work is the most minimal model of radiative Dirac neutrino mass \cite{Saad:2019bqf} and the second model uses the scotogenic mechanism \cite{Ma:2006km} that links two seemingly uncorrelated phenomena:  neutrino mass with Dark Matter (DM). As we will discuss, the stability of the DM in the latter scenario is a consequence of a residual $\mathcal{Z}_2$ discrete symmetry that emerges from the spontaneous breaking of the $U(1)_R$ gauge symmetry.   

Among other simple possibilities,  one can also extend  the SM with  $U(1)_{B-L}$ gauge symmetry \cite{BminusL} for generating the Dirac neutrino mass \cite{Wang:2017mcy, Han:2018zcn, Calle:2018ovc, Bonilla:2018ynb, Saad:2019bqf, Bonilla:2019hfb}. Both of the two possibilities are attractive and  can be regarded as the minimal gauge extensions of the SM. However, the phenomenology of $U(1)_R$ model is very distinctive compared to the $U(1)_{B-L}$ case. In the literature, gauged  $U(1)_{B-L}$ symmetry has been extensively studied  whereas gauged $U(1)_R$ extension has received very little  attention. 

Unlike the $U(1)_{B-L}$ case, in our set-up, the SM Higgs doublet is  charged under this $U(1)_R$ symmetry to allow the desired Yukawa interactions to generate mass for the charged fermions,  this leads to interactions with the new gauge boson that is absent in $U(1)_{B-L}$ model. The running of the Higgs quartic coupling gets modified  due to having such interactions with the new gauge boson $Z^{\prime}$ that can make the Higgs vacuum stable \cite{Chao:2012mx}.  Due to the same reason, the SM Higgs phenomenology also gets altered \cite{Ko:2012hd}.  

We show by detail analysis that despite  their abelian nature, $U(1)_{R}$ and $U(1)_{B-L}$ have distinguishable phenomenology. The primary reason that leads to different features is:  $U(1)_R$ gauge boson couples only to the right-handed chiral fermions, whereas $U(1)_{B-L}$ is chirality-universal. As a consequence, $U(1)_R$ model leads to large left-right (LR) asymmetry and also forward-backward (FB) asymmetry that can be tested in the current and future colliders that make use of the polarized initial states, such as in ILC.  We also comment on the differences of our $U(1)_R$ scenario with the other $U(1)_R$ models existing in the literature. Slightly different features emerge as a result of different charge assignment of the right-handed neutrinos in our set-up for the realization of Dirac neutrino mass. In the existing $U(1)_R$ models, flavor universal charge assignment for the right-handed neutrinos are considered and neutrinos are assumed to be Majorana particles. Whereas, in our set-up, neutrinos are Dirac particles that demands non-universal charge assignment for the right-handed neutrinos under $U(1)_R$.  Neutrinos being Dirac in nature also leads to null neutrinoless double beta decay signal.  

The originality of this work is,  by  employing only the gauged  $U(1)_R$ symmetry, we construct Dirac neutrino masses at the tree-level and one-loop level (with or without DM) which has not been done before and,  by a detailed  study of the phenomenology associated  to the new heavy gauge boson, we show that $U(1)_R$ model is very promising to be discovered in the future colliders. Due to the presence of the TeV or sub-TeV scale BSM particles, these models can give rise to sizable rate for the charged lepton flavor violating processes which we also analyze.  On top of that, we bring both the dark matter and the neutrino mass generation issues under one umbrella without imposing any additional symmetry and, work out the  associated dark matter phenomenology. We  also discuss  the cosmological consequences due to the presence of the light right-handed neutrinos in our framework. 

The paper is organized as follows.  In Section \ref{SEC-02}, we discuss the framework where SM is extended by an abelian gauge symmetry $U(1)_R$. In Section \ref{sec:dirac}, we present the minimal Dirac neutrino mass models in details, along with the particle spectrum and charge assignments. In Section \ref{running_gaugec}, we discuss the running of the $U(1)_R$ coupling. Charged lepton flavor violating processes are analyzed in Section \ref{SEC-LFV}. We have also done the associated dark matter phenomenology in Section \ref{SEC-DM} for the scotogenic model. Furthermore, we analyze the collider implications in Section \ref{sec:collider}. In Section \ref{cosmo}, we study the constraints from cosmological measurement and finally,  we conclude in Section \ref{con}.

\section{Framework}\label{SEC-02} 
Our framework is a very simple extension of the SM: an abelian gauge extension under which only the right-handed fermions are charged. Such a charge assignment is anomalous, however, all the gauge anomalies can be canceled by the minimal extension of the SM with just three right-handed neutrinos.   Within this framework the minimal choice to generate the charged fermion masses is to utilize the already existing  SM Higgs doublet, hence the associated Yukawa couplings have the form:
\begin{align}
\mathcal{L}_Y\supset
y_u\overline{Q}_L\widetilde{H}u_R
+
y_d\overline{Q}_L H d_R
+
y_e\overline{L}_L H \ell_R
+ h.c.
\end{align}

\noindent 
As a result, the choice of the $U(1)_R$ charges of the right-handed fermions of the SM must be universal and  obey the following relationship:
\begin{align}
R_u=-R_d=-R_{\ell}=R_H.    
\end{align}

\noindent
Here $R_k$ represents the $U(1)_R$ charge of the particle $k$. 
Hence,  all the charges are determined  once $R_H$ is fixed, which can take any value. The anomaly is canceled by the presence of the  right-handed neutrinos that in general can carry non-universal charge under $U(1)_R$.  Under the symmetry of the theory, the  quantum numbers of all the particles are shown in Table \ref{charge-SM}.

\FloatBarrier
\begin{table}[t!]
\centering
\footnotesize
\resizebox{0.5\textwidth}{!}{
\begin{tabular}{|c|c|}
\hline
Multiplets& $SU(3)_C\times SU(2)_L\times U(1)_Y\times U(1)_{R}$   \\ \hline\hline
Quarks&
\pbox{10cm}{
\vspace{2pt}
${Q_L}_i (3,2,\frac{1}{6},\textcolor{blue}{0})$\\
${u_R}_i (3,1,\frac{2}{3},\textcolor{blue}{R_H})$\\
${d_R}_i (3,1,-\frac{1}{3},\{\textcolor{blue}{-R_H}\})$
\vspace{2pt}}
\\ \hline\hline
Leptons&
\pbox{10cm}{
\vspace{2pt}
${L_L}_i (1,2,-\frac{1}{2},\textcolor{blue}{0})$\\
${\ell_R}_i (1,1,-1,\textcolor{blue}{-R_H})$\\
${\nu_R}_i (1,1,0,\{\textcolor{blue}{R_{\nu_1},R_{\nu_2},R_{\nu_3}}\})$
\vspace{2pt}}
\\ \hline\hline
Higgs &
\pbox{10cm}{
\vspace{2pt}
$H (1,2,\frac{1}{2},\textcolor{blue}{R_H})$
\vspace{2pt}}
\\ \hline
\end{tabular}
}
\caption{ 
Quantum numbers of the fermions and the SM Higgs doublet.
}\label{charge-SM}
\end{table}

In our set-up, all the anomalies automatically cancel except for the following two:
\begin{align}
&[U(1)_R]:  R_{\nu_1}+R_{\nu_2}+R_{\nu_3}=3R_H,   
\\
&[U(1)_R]^3:  R_{\nu_1}^3+R_{\nu_2}^3+R_{\nu_3}^3=3R_H^3. 
\end{align}

\noindent
This system has two different types of solutions. 
The simplest solution corresponds to the case  of flavor universal charge assignment that demands: $R_{\nu{1,2,3}}=R_H$ which has been studied in the literature \cite{Nomura:2016pgg, Nomura:2017ezy, Nomura:2017tih, Chao:2017rwv,  Nomura:2018mwr}. In this work, we adopt the alternative choice of flavor non-universal solution and show that the predictions and phenomenology of this set-up can be very  different from the flavor universal scenario.  We compare our model with the other $U(1)_R$ extensions, as well as $U(1)_{B-L}$ extensions of the SM. As already pointed out, a different charge assignment leads to distinct phenomenology  in our model and can be distinguished in  the neutrino and collider experiments.  

Since SM is a good symmetry at the low energies, $U(1)_R$ symmetry needs to be broken around $O(10)$ TeV scale or above. We assume that $U(1)_R$ gets broken spontaneously by the VEV of a SM singlet $\chi (1,1,0,R_{\chi})$ that must carry non-zero charge ($R_{\chi}\neq 0$) under $U(1)_R$. As a result of this symmetry breaking, the imaginary part of $\chi$ will be eaten up by the corresponding gauge boson $X_{\mu}$ to become  massive.  Since EW symmetry also needs to break down around the $O(100)$ GeV scale, one can compute the  masses of the gauge bosons from the covariant derivatives associated with the SM Higgs $H$ and the SM singlet scalar $\chi$:
\begin{align}
&D_{\mu}H=\left( \partial_{\mu} -i g W_{\mu}-i g^{\prime} Y_H B_{\mu} - i g_R R_H X_{\mu} \right)H,
\\
&D_{\mu}\chi=\left( \partial_{\mu} - i g_R R_{\chi} X_{\mu} \right)\chi.
\end{align}

\noindent 
As a consequence of the symmetry breaking, the neutral components of the gauge bosons will all mix with each other. 
Inserting the following VEVs:
\begin{align}
\langle H \rangle= \begin{pmatrix}
0\\\frac{v_H}{\sqrt{2}}
\end{pmatrix},\;\;\; \langle \chi \rangle= \frac{v_{\chi}}{\sqrt{2}},
\end{align}

\noindent
one can compute the neutral gauge boson masses as:
\begin{align}\label{Zmatrix}
\begin{pmatrix}
B&W_3&X
\end{pmatrix}
\left( \frac{v^2_H}{4} \right)
\begin{pmatrix}
g^{\prime 2}&-g^{\prime}g&2g^{\prime}g_RR_H\\
-g^{\prime}g&g^2&-2g g_R R_{\chi}\\
2g^{\prime}g_RR_H&-2g g_R R_{\chi}&4g^2_RR_H^2(1+r_v^2)
\end{pmatrix}
\begin{pmatrix}
B\\W_3\\X
\end{pmatrix}.
\end{align}

\noindent Where, $r_v= \frac{R_{\chi}v_{\chi}}{R_{H}v_{H}}$ and the well-known relation $\tan\theta_w=g^{\prime}/g$ and furthermore $v_H=246$ GeV. In the above mass matrix denoted by  $M^2$,  one of the gauge bosons remains massless, which must be  identified as the photon field, $A_{\mu}$. Moreover, two massive states appear which  are the SM $Z$-boson and a heavy $Z^{\prime}$-boson ($M_{Z}<M_{Z^{\prime}}$). The corresponding masses are given by:
\begin{align}\label{Zmass}
M_{Z,Z^{\prime}}= \frac{gv_H}{2c_w} \left( 
\frac{1}{2}\left[ 1+r^2_Xc^2_w(1+r^2_v)  \right]
\mp \left[  \frac{r_Xc_w}{\sin(2\theta_X)} \right]
\right)^{\frac{1}{2}},
\end{align}

\noindent 
here we define:
\begin{align}
&r_X= (2g_RR_H)/g,\\
&\sin(2\theta_X)= \frac{2r_Xc_w}{\left( 
\left[ 2r_Xc_w \right]^2 + 
\left[ (1+r^2_v)r^2_Xc^2_w-1 \right]^2
\right)^{\frac{1}{2}}}.
\end{align}

\noindent
Which clearly shows that for $g_R=0$, the mass of the SM gauge boson is reproduced: $M^{SM}_Z= \frac{1}{2}v_H(g^2+g^{\prime 2})^{1/2}= \frac{1}{2} gv_H/c_w$. 
To find the corresponding eigenstates, we diagonalize the mass matrix as: $M^2= U^{\dagger} M^2_{diag}U^{\ast}$, with:
\begin{align}
\begin{pmatrix}
B\\W_3\\X
\end{pmatrix}
=U
\begin{pmatrix}
A\\Z\\Z^{\prime}
\end{pmatrix},
\;\;\;
U=\begin{pmatrix}
c_w&-s_wc_X&s_ws_X\\
s_w&c_wc_X&-c_ws_X\\
0&s_X&c_X
\end{pmatrix}.    
\end{align}

\noindent
From Eq. \eqref{Zmass} one can see that the mass of  the SM $Z$-boson gets modified as a consequence of $U(1)_R$ gauge extension. Precision measurement of the SM $Z$-boson puts bound on the scale of the new physics. From the experimental measurements, the bound on the lower limit of the new physics scale can be found by imposing the constraint  $\Delta M_{Z} \leq 2.1$ MeV \cite{Tanabashi:2018oca}. For our case, this bound can be translated into: 
\begin{align}
\left|\Delta M_Z\right|  = \left| M^{SM}_Z \left(  1- \sqrt{\frac{r^2_v}{1+r^2_v}} \right)\right|   \leq 2.1 \;\rm{MeV}. 
\end{align}

\noindent With $M^{SM}_Z=91.1876$ GeV \cite{Tanabashi:2018oca}, we find $v_{\chi}\geq \left( \frac{v_HR_H}{R_{\chi}} \right) 21708.8$. Which  corresponds to $v_{\chi}\geq 12.08$ TeV for $R_H=1$ and $R_{\chi}=3$ (this charge assignment for the SM Higgs doublet $H$ and the SM singlet scalar $\chi$ that breaks $U(1)_R$ will be used in Secs. \ref{sec:dirac} and \ref{sec:collider}).

Furthermore,  the coupling of all the fermions with the new gauge boson can be computed from the following relevant part of the Lagrangian:
\begin{align}
\mathcal{L}\supset 
g_{\psi}\;\overline{\psi}\gamma^{\mu}Z^{\prime}_{\mu}\psi.
\end{align}

\noindent 
The couplings $g_{\psi}$ of all the fermions in our theory are collected in Table \ref{gauge} and will be useful for our  phenomenological study performed later in the text. 
Note that the couplings of the left-handed SM fermions are largely suppressed compared to the right-handed  ones, since they are always proportional to $\sin \theta_X$ and $\theta_X$ must be small and is highly constrained  by the experimental data.   

\FloatBarrier
\begin{table}[t!]
\centering
\footnotesize
\resizebox{0.7\textwidth}{!}{
\begin{tabular}{|c|c|}
\hline
Fermion, $\psi$& Coupling, $g_{\psi}$   \\ \hline\hline
Quarks&
\pbox{10cm}{
\vspace{2pt}
$g_{u_L}=-\frac{1}{6}\frac{g}{c_w}(1+2c_{2w})s_X$\\
$g_{d_L}=\frac{1}{6}\frac{g}{c_w}(2+c_{2w})s_X$\\
$g_{u_R}=\frac{2}{3}\frac{g}{c_w}s^2_ws_X+ g_Rc_XR_H$\\
$g_{d_R}=-\frac{1}{3}\frac{g}{c_w}s^2_ws_X- g_Rc_XR_H$
\vspace{2pt}}
\\ \hline\hline
Leptons&
\pbox{10cm}{
\vspace{2pt}
$g_{\nu_L}=-\frac{1}{2}\frac{g}{c_w}s_X$\\
$g_{\ell_L}=\frac{1}{2}\frac{g}{c_w}c_{2w}s_X$\\
$g_{\ell_R}=-\frac{g}{c_w}s^2_ws_X- g_Rc_XR_H$\\
$g_{\nu_{R_i}}=g_Rc_XR_{\nu_i}$
\vspace{2pt}}
\\ \hline\hline
Vector-like fermions &
\pbox{10cm}{
\vspace{2pt}
$g_{\mathcal{N}}=g_Rc_XR_{\mathcal{N}}$
\vspace{2pt}}
\\ \hline
\end{tabular}
}
\caption{ 
Couplings of the fermions with the new gauge boson. Here we use the notation: $c_{2w}=\cos(2\theta_w)$. $\mathcal{N}_{L,R}$ is any vector-like fermion singlet under the SM and carries  $R_{\mathcal{N}}$ charge under  $U(1)_R$. If a model does not contain vector-like fermions, we set $R_{\mathcal{N}}=0$. 
}\label{gauge}
\end{table}

Based on the framework introduced in this section, we construct various minimal  models of Dirac neutrino masses in Sec \ref{sec:dirac} and study various phenomenology in the subsequent sections.

\section{Dirac Neutrino Mass Models}\label{sec:dirac}
By adopting the set-up as discussed above in this section, we construct models of Dirac neutrino masses. Within this set-up, 
if the solution $R_{\nu_i}=R_H$ is chosen which is allowed by the anomaly cancellation conditions, then tree-level Dirac mass term $y_{\nu}v_H \overline{\nu}_L\nu_R$ is allowed and observed oscillation data requires tiny Yukawa couplings of order $y_{\nu}\sim 10^{-11}$. This is expected not to be a natural scenario, hence due to aesthetic reason we generate naturally small Dirac neutrino mass by exploiting the already existing symmetries in the theory. This requires the  implementation of  the flavor non-universal solution of the anomaly cancellation conditions, in such a scenario   $U(1)_R$ symmetry plays the vital role in forbidding the direct Dirac mass term and also all Majorana mass terms for the neutrinos.

In this section, we explore three different models within our framework  where neutrinos receive naturally small Dirac mass either at the tree-level or at the one-loop level. Furthermore, we also show that the stability of DM can be assured by a residual discrete symmetry resulting from the spontaneous symmetry breaking of $U(1)_R$. In the literature,  utilizing $U(1)_R$ symmetry,  two-loop Majorana neutrino mass is constructed with the imposition of an additional $\mathcal{Z}_2$ symmetry in  \cite{Nomura:2016pgg, Nomura:2017ezy} and three types of seesaw cases are discussed,  standard type-I seesaw in \cite{Nomura:2017tih}, type-II seesaw in \cite{Chao:2017rwv} and   inverse seesaw model in \cite{Nomura:2018mwr}. In constructing the inverse seesaw model, in addition to $U(1)_R$,  additional flavor dependent U(1) symmetries are also imposed in \cite{Nomura:2018mwr}. In all these models, neutrinos are assumed to be Majorana particles which is not the case in our scenario.

\subsection{Tree-level Dirac Seesaw}
\begin{figure}[th!]
\centering
$$
\includegraphics[height=6cm,width=0.7\textwidth]{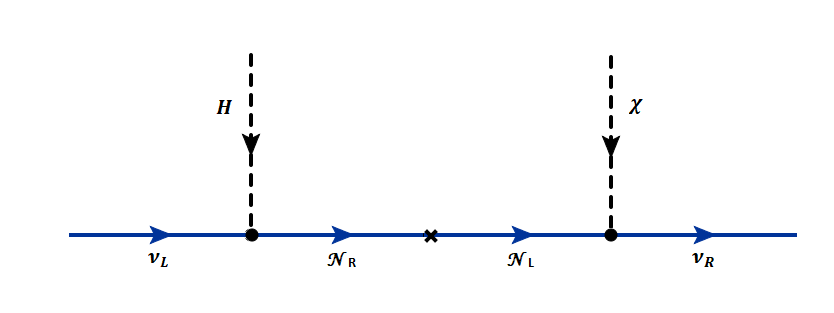}\vspace{-0.5in}
 $$
\caption{Representative Feynman diagram for tree-level Dirac Seesaw. }\label{R1}
\end{figure}

In this sub-section, we focus on the tree-level neutrino mass generation via Dirac seesaw mechanism \cite{Roy:1983be}\footnote{For correlating Dirac seesaw with leptogenesis, see for example \cite{Gu:2012fg, Gu:2016hxh}.}. For the realization of this scenario, we introduce three generations of vector-like fermions that are singlets under the SM:  $\mathcal{N}_{L,R}(1,1,0,R_{\mathcal{N}})$. In this model,  
the quantum numbers of the multiplets are shown in Table \ref{model-1} and the corresponding Feynman diagram for neutrino mass generation is shown in Fig. \ref{R1}.
 This choice of the particle content allows one to write the following Yukawa coupling terms relevant for neutrino mass generation:
\begin{align}
\mathcal{L}_Y\supset 
y^H\overline{L}_L\widetilde{H}\mathcal{N}_R
+M_{\mathcal{N}}\overline{\mathcal{N}}_L\mathcal{N}_R
+y^{\chi}\overline{\mathcal{N}}_L\nu_R\chi^{\ast} 
+h.c.
\end{align}

\noindent
Here, we have suppressed the generation and the group indices. And the Higgs potential is given by:
\begin{equation} 
    V= -\mu^2_HH^{\dagger}H+\lambda (H^{\dagger}H)^2-\mu_{\chi}^2\chi^{*}\chi +\lambda_1(\chi^{*}\chi)^2+\lambda_2H^{\dagger}H\chi^{*}\chi.
\end{equation}

\noindent 
When both the $U(1)_R$ and EW symmetries are broken, the part of the above Lagrangian responsible for neutrino mass generation can be written as:
\begin{align}
\mathcal{L}_Y\supset  
\begin{pmatrix}
\overline{\nu}_L&\overline{\mathcal{N}}_L
\end{pmatrix}
M_{\nu,\mathcal{N}}
\begin{pmatrix}
\nu_R\\\mathcal{N}_R
\end{pmatrix},
\;\;\;
M_{\nu,\mathcal{N}}=
\begin{pmatrix}
0&\frac{v_H}{\sqrt{2}}y^H\\
\frac{v_{\chi}}{\sqrt{2}}y^{\chi}&M_{\mathcal{N}}
\end{pmatrix}.
\end{align}

\noindent
Where $M_{\nu,\mathcal{N}}$ is a $6\times 6$ matrix and, since $\nu_{R_1}$ carries a different charge we have $y^{\chi}_{i1}=0$.  The bare mass term $M_{\mathcal{N}}$ of the vector-like fermions can in principle be large compared to the two VEVs, $M_{\mathcal{N}}\gg v_{H,\chi}$, assuming this scenario the light neutrino masses are given by:
\begin{align}
m_{\nu}\sim 
\frac{v_Hv_{\chi}}{2} 
\frac{y^Hy^{\chi}}{M_{\mathcal{N}}}.
\end{align}
 
\noindent 
Assuming $v_{\chi}=10$ TeV, $y^H=y^{\chi}\sim 10^{-3}$, to get $m_{\nu}=0.1$ eV one requires $M_{\mathcal{N}}\sim 10^{10}$ GeV. Dirac neutrino mass generation of this type from a generic point of view without specifying the underline symmetry is discussed in 
\cite{Ma:2016mwh}.

\begin{table}[th!]
\centering
\footnotesize
\resizebox{0.6\textwidth}{!}{
\begin{tabular}{|c|c|}
\hline
Multiplets& $SU(3)_C\times SU(2)_L\times U(1)_Y\times U(1)_{R}$   \\ \hline\hline
Leptons&
\pbox{10cm}{
\vspace{2pt}
${L_L}_i (1,2,-\frac{1}{2},\textcolor{blue}{0})$\\
${\ell_R}_i (1,1,-1,\textcolor{blue}{-1})$\\
${\nu_R}_i (1,1,0,\{\textcolor{blue}{-5, 4, 4}\})$
\vspace{2pt}}
\\ \hline\hline
Scalars &
\pbox{10cm}{
\vspace{2pt}
$H (1,2,\frac{1}{2},\textcolor{blue}{1})$\\
$\chi (1,1,0,\textcolor{blue}{3})$
\vspace{2pt}}
\\ \hline \hline 
Vector-like fermion&
\pbox{10cm}{
\vspace{2pt}
$\mathcal{N}_{L,R} (1,1,0,\textcolor{blue}{1})$
\vspace{2pt}}
\\ \hline
\end{tabular}
}
\caption{ 
Quantum numbers of the fermions and the scalars  in Dirac seesaw model.
}\label{model-1}
\end{table}

In this scenario  two chiral massless states appear, one of them  is $\nu_{R_1}$,  which is a consequence of its charge being different from the other two generations.  In principle,  all three generations of neutrinos can be given Dirac mass if the model is extended by a second SM singlet $\chi^{\prime}(1,1,0,-6)$. When this field  acquires an induced  VEV all neutrinos become massive. This new SM singlet scalar, if introduced,  gets an induced VEV from a cubic coupling of the form: $\mu \chi^2 \chi^{\prime} +h.c.$.    Alternatively, without specifying the ultraviolet completion of the model, a small Dirac neutrino mass for the massless chiral states can be generated via the dimension-5 operator $\overline{\mathcal{N}}_L\nu_R\langle \chi\rangle \langle \chi \rangle /\Lambda$   once $U(1)_R$ is broken spontaneously.

\subsection{Simplest one-loop implementation}
In this sub-section, we consider the most minimal \cite{Saad:2019bqf} model of radiative Dirac neutrino mass  in the context of $U(1)_R$ symmetry.  Unlike the previous sub-section, we do not introduce any vector-like fermions, hence neutrino mass does not appear at the tree-level. All tree-level Dirac and Majorana neutrino mass terms are automatically forbidden due to $U(1)_R$ symmetry reasons. This model consists of two singly charged scalars $S^+_i$ to complete the loop-diagram and a neutral scalar $\chi$ to break the $U(1)_R$ symmetry,   the  particle content with their quantum numbers is presented in Table \ref{model-2}. 

\begin{table}[th!]
\centering
\footnotesize
\resizebox{0.55\textwidth}{!}{
\begin{tabular}{|c|c|}
\hline
Multiplets& $SU(3)_C\times SU(2)_L\times U(1)_Y\times U(1)_{R}$   \\ \hline\hline
Leptons&
\pbox{10cm}{
\vspace{2pt}
${L_L}_i (1,2,-\frac{1}{2},\textcolor{blue}{0})$\\
${\ell_R}_i (1,1,-1,\textcolor{blue}{-1})$\\
${\nu_R}_i (1,1,0,\{\textcolor{blue}{-5, 4, 4}\})$
\vspace{2pt}}
\\ \hline\hline
Scalars &
\pbox{10cm}{
\vspace{2pt}
$H (1,2,\frac{1}{2},\textcolor{blue}{1})$\\
$\chi (1,1,0,\textcolor{blue}{3})$\\
$S^+_1 (1,1,1,\textcolor{blue}{0})$\\
$S^+_2 (1,1,1,\textcolor{blue}{-3})$
\vspace{2pt}}
\\ \hline 
\end{tabular}
}
\caption{ 
Quantum numbers of the fermions and the scalars  in radiative Dirac  model.
}\label{model-2}
\end{table}

With this particle content, the gauge invariant terms in the Yukawa sector responsible for generating neutrino mass are  given by:
\begin{align}
\mathcal{L}_Y\supset
y^H\overline{L}_L\ell_R H
+y^{S_1}\overline{L^c_L}\epsilon L_L S^+_1
+y^{S_2}\overline{\nu^c_R} \ell_R S^+_2
+h.c.
\end{align}

\noindent
And the complete Higgs potential is given by:
\begin{align}
    V &= -\mu_H^2H^{\dagger}H+\mu_1^2|S_1^+|^2+\mu_2^2|S_2^+|^2-\mu_{\chi}^2\chi^{*}\chi +(\mu S_2^+S_1^-\chi+h.c.)+\lambda (H^{\dagger}H)^2
    +\lambda_1|S_1^+|^4 
 + \lambda_2|S_2^+|^4
   \nonumber \\ &
 +\lambda_{\chi}(\chi^{*}\chi)^2+\lambda_3|S_1^+|^2|S_2^+|^2+\lambda_4|S_1^+|^2H^{\dagger}H+\lambda_5|S_2^+|^2H^{\dagger}H+\lambda_6H^{\dagger}H\chi^{*}\chi.
\end{align}

\begin{figure}[th!]
\centering
$$
\includegraphics[height=7.5cm,width=0.75\textwidth]{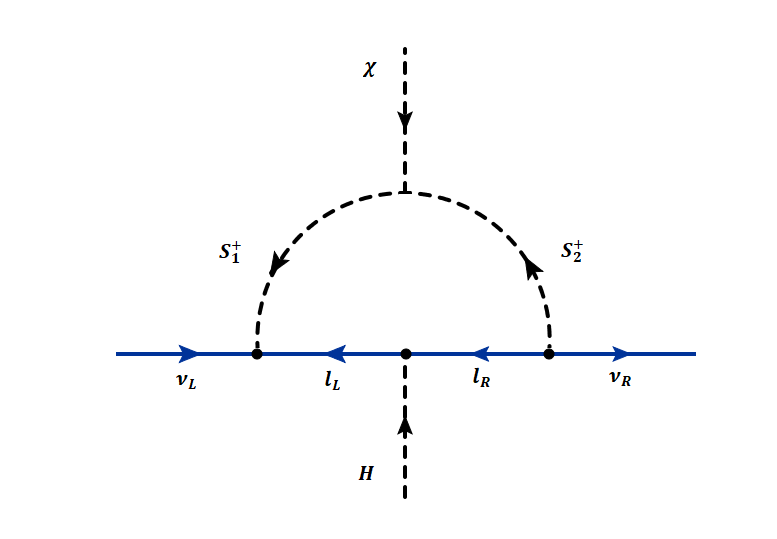} \vspace{-0.5in}
 $$
\caption{Representative Feynman diagram for the simplest one-loop Dirac neutrino mass. 
}\label{R2}
\end{figure}

\noindent
By making use of the existing  cubic term $V\supset \mu S_2^+S_1^-\chi+h.c.$ one can draw the desired one-loop Feynman diagram that is  presented in Fig. \ref{R2}. The neutrino mass matrix in this model is given by:
\begin{align}
{m_{\nu}}_{ab}= \frac{\sin(2\theta)}{16\pi^2}\text{ln}\left( \frac{m^2_{H_2}}{m^2_{H_1}} \right) y^{S_1}_{ai}{m_E}_{i}y^{S_2}_{ib}.    
\end{align}

\noindent
Here $\theta$ represents the mixing between the singly charged scalars and $m_{H_i}$  represents the mass of the physical state $H^+_i$.  Here we make a crude estimation of the neutrino masses: 
for $\theta=0.1$ radian, $m_{H_2}/m_{H_1}=1.1$ and  $y^{S_i}\sim 10^{-3}$ one gets the correct order of neutrino mass $m_{\nu}=0.1$ eV.

This is the most minimal radiative Dirac neutrino mass mechanism which was constructed by employing a $\mathcal{Z}_2$ symmetry in
\cite{Nasri:2001ax} and just recently in  \cite{Calle:2018ovc, Saad:2019bqf} by utilizing $U(1)_{B-L}$ symmetry. As a result of the anti-symmetric property of the Yukawa couplings  $y^{S_1}$, one pair of chiral states remains massless to all orders, higher dimensional operators cannot induce mass to all the neutrinos. As already pointed out, neutrino oscillation data is not in conflict with one massless state.

\subsection{Scotogenic Dirac neutrino mass}
\begin{figure}[th!]
\centering
$$
\includegraphics[height=7cm,width=0.7\textwidth]{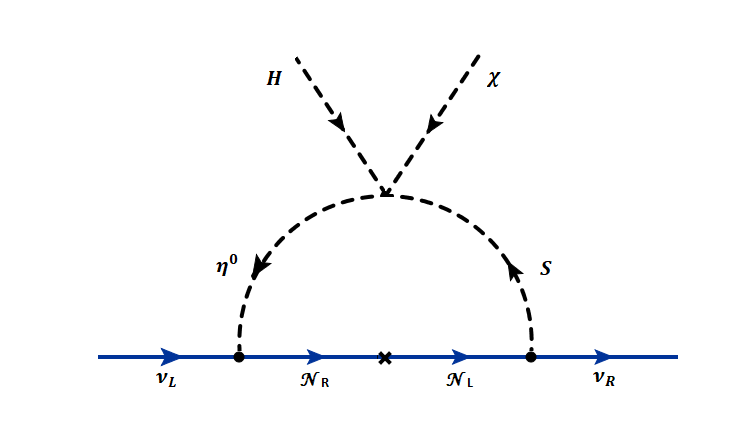} \vspace{-0.5in}
 $$
\caption{Representative Feynman diagram for scotogenic Dirac neutrino mass model. 
}\label{R3}
\end{figure}

The third possibility of Dirac neutrino mass generation that we discuss in this sub-section contains a DM candidate. The model we present here   belongs to the radiative scotogenic \cite{Ma:2006km} class of models  
and contains a second Higgs doublet in addition to two SM singlets. Furthermore, a vector-like fermion singlet under the SM is required to complete the one-loop diagram. The particle content of this model is listed in Table \ref{model-3} and the associated loop-diagram is presented in Fig. \ref{R3}. 

\begin{table}[th!]
\centering
\footnotesize
\resizebox{0.6\textwidth}{!}{
\begin{tabular}{|c|c|c|}
\hline
Multiplets& $SU(3)_C\times SU(2)_L\times U(1)_Y\times U(1)_{R}$  \\ \hline\hline
Leptons&
\pbox{10cm}{
\vspace{2pt}
${L_L}_i (1,2,-\frac{1}{2},\textcolor{blue}{0})$\\
${\ell_R}_i (1,1,-1,\textcolor{blue}{-1})$\\
${\nu_R}_i (1,1,0,\{\textcolor{blue}{-5, 4, 4}\})$
\vspace{2pt}}
\\ \hline\hline
Scalars &
\pbox{10cm}{
\vspace{2pt}
$H (1,2,\frac{1}{2},\textcolor{blue}{1})$\\
$\chi (1,1,0,\textcolor{blue}{3})$\\
$S (1,1,0,\textcolor{blue}{-\frac{7}{2}})$\\
$\eta (1,2,\frac{1}{2},\textcolor{blue}{\frac{1}{2}})$
\vspace{2pt}}
\\ \hline\hline 
Vector-like fermion&
\pbox{10cm}{
\vspace{2pt}
$\mathcal{N}_{L,R} (1,1,0,\textcolor{blue}{\frac{1}{2}})$
\vspace{2pt}}
\\ \hline 
\end{tabular}
}
\caption{ 
Quantum numbers of the fermions and the scalars  in scotogenic Dirac neutrino mass model.  
}\label{model-3}
\end{table}

The relevant Yukawa interactions are given as follows:
\begin{align}
y^{\eta}\overline{L}_L\mathcal{N}_R\widetilde{\eta} 
+M_{\mathcal{N}} \overline{\mathcal{N}}_L\mathcal{N}_R
+y^{S}\overline{\mathcal{N}}_L\nu_RS+h.c.
\end{align}

\noindent
And the complete Higgs potential is given by:
\begin{align}
    V &= -\mu^2_H H^{\dagger}H+\lambda (H^{\dagger}H)^2+\mu^2_{\eta}\eta^{\dagger}\eta+ \lambda_{\eta}(\eta^{\dagger}\eta)^2-\mu^2_{\chi}\chi^*\chi+\lambda_{\chi}(\chi^*\chi)^2+\mu^2_SS^*S+\lambda_S(S^*S)^2
    \nonumber \\ &
   +\lambda_1H^{\dagger}H\eta^{\dagger}\eta +\lambda_2H^{\dagger}HS^*S
   +\lambda_3H^{\dagger}H\chi^*\chi+\lambda_4\eta^{\dagger}\eta S^*S+\lambda_5\eta^{\dagger}\eta\chi^*\chi
    +\lambda_6\chi^*\chi S^*S
    \nonumber \\ &
    +(\lambda_7 H^{\dagger}\eta \eta^{\dagger}H+h.c.)
    +(\lambda_D\eta^{\dagger}H\chi S+h.c.).
\end{align}

The SM singlet $S$ and the second Higgs doublet $\eta$ do not acquire any VEV and the loop-diagram is completed by making use of the quartic coupling $V\supset \lambda_D\eta^{\dagger}H\chi S+h.c.$.
Here for simplicity, we assume that the SM Higgs does not mix with the other CP-even states, consequently,  the mixing between $S^0$ and $\eta^0$ originates from the quartic coupling  $\lambda_{D}$  (and similarly for the CP-odd states). Then the neutrino mass matrix is given by:
\begin{align}
{m_{\nu}}_{ab}&=
\frac{1}{16\pi^2}\frac{\sin\theta\cos\theta}{2}
{y^{\eta}}_{ai}{M_{\mathcal{N}}}_{i}{y^S}_{ib}
\left(  F\left[ \frac{m^2_{H^0_2}}{{M^2_{\mathcal{N}}}_{i}} \right] - F\left[ \frac{m^2_{H^0_1}}{{M^2_{\mathcal{N}}}_{i}} \right] \right)
\\&-
\frac{1}{16\pi^2}\frac{\sin\theta^{\prime}\cos\theta^{\prime}}{2}
{y^{\eta}}_{ai}{M_{\mathcal{N}}}_{i}{y^S}_{ib}
\left(  F\left[ \frac{m^2_{A^0_2}}{{M^2_{\mathcal{N}}}_{i}} \right] - F\left[ \frac{m^2_{A^0_1}}{{M^2_{\mathcal{N}}}_{i}} \right] \right).
\end{align}

\noindent
Where the mixing angle $\theta$ ( $\theta^{\prime}$) between the CP-even (CP-odd) states are given by:
\begin{align}
&\theta=\frac{1}{2}\sin^{-1}\left( \frac{\lambda_{D}\;v_H\;v_{\chi}}{m^2_{H^0_2}-m^2_{H^0_1}} \right),
\;\;\;
\theta^{\prime}=\frac{1}{2}\sin^{-1}\left( \frac{\lambda_{D}\;v_H\;v_{\chi}}{m^2_{A^0_2}-m^2_{A^0_1}} \right). 
\end{align}

\noindent
For a rough estimation we assume no cancellation among different terms occurs. Then by setting  $m_H=1$ TeV, $M_{\mathcal{N}}=10^3$ TeV, $\lambda_D=0.1$, $v_{\chi}=10$ TeV, $y^{\eta,S}\sim 10^{-3}$ one can get the correct order of neutrino mass  $m_{\nu}\sim 0.1$ eV. 

Since $\nu_{R_1}$ carries a charge of $-5$, a pair of chiral  states associated with this state remains massless. However, in this scotogenic version, unlike the simplest one-loop model presented in the previous sub-section, all the neutrinos can be given mass by extending the model further. Here just for completeness, we discuss a straightforward extension, even though this is not required since one massless neutrino is not in conflict with the experimental data. If the  model defined by Table \ref{model-3}   is extended by two SM singlets $\chi^{\prime}(1,1,0,-6)$ and a $S^{\prime}(1,1,0,\frac{11}{2})$, all the neutrinos will get non-zero mass. The  VEV of the field $\chi^{\prime}$ can be induced by the allowed cubic term of the form $\mu \chi^2\chi^{\prime}+h.c.$ whereas, $S^{\prime}$ does not get any induced VEV.   
 
 Here we comment on the DM candidate present in this model. As aforementioned, we do not introduce new symmetries by hand to stabilize the DM. In search of  finding the  unbroken symmetry, first, we rescale all the $U(1)_R$ charges of the particles in the theory given in Table \ref{model-3} including the quark fields in such a way that the magnitude of the minimum charge is unity. From this rescaling, it is obvious that when the $U(1)_R$ symmetry is broken spontaneously by the VEV of the $\chi$ field that carries six units of rescaled charge leads to: $U(1)_R\to \mathcal{Z}_6$.    However, since the SM Higgs doublet carries a charge of two units under this surviving $\mathcal{Z}_6$ symmetry, its VEV further breaks this symmetry down to: $\mathcal{Z}_6\to \mathcal{Z}_2$.  This unbroken discrete $\mathcal{Z}_2$ symmetry can stabilize the DM particle in our theory. Under this residual  symmetry, all the SM particles are even, whereas only the scalars $S, \eta$ and vector-like fermions $\mathcal{N}_{L,R}$ are odd and can be the DM candidate. Phenomenology  associated with the DM matter in this scotogenic model will be discussed in Sec. \ref{SEC-DM}.

\section{Running of the $U(1)_R$ Gauge Coupling} \label{running_gaugec}
In this section, we briefly discuss the running of the $U(1)_R$ gauge coupling $g_R$, at the one-loop level in our framework.  The associated $\beta$-function can be written as:
\begin{align}\label{running}
\beta_R =\frac{1}{16 \pi^2} b_R g^3_R.   
\end{align}
Where the coefficient $b_R$ can be calculated from  \cite{Machacek:1983tz}:
\begin{align}
b_R=\sum_{f_i} \frac{4}{3}\kappa N_g S_2(f_i) + \sum_{s_i} \frac{1}{6} \eta S_2(s_i).    
\end{align}

\FloatBarrier
\begin{figure}[b!]
\centering
\includegraphics[scale=0.5]{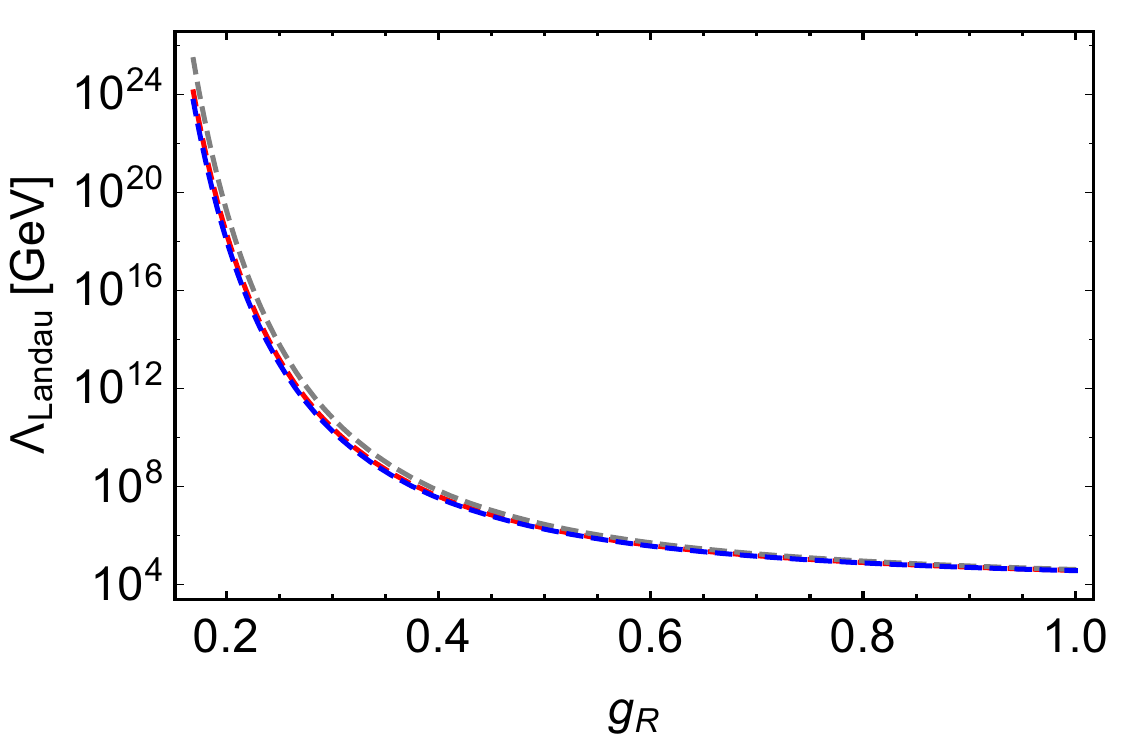} 
\caption{Possible presence of Landau poles associated with $U(1)_R$ gauge coupling running. For this plot, we have fixed $\mu_0=10$ TeV. Red, gray and blue lines correspond to Dirac seesaw, simplest one-loop and Scotogenic models respectively. 
}\label{U1-running}
\end{figure}

\noindent
The first (second) sum is over the fermions (scalars), $f_i$ ($s_i$). Here, $\kappa=1/2$ for Weyl fermions, $N_g$ is the number of fermion generations, $\eta=2$ for complex scalars and $S_2$ are the Dynkin indices of the representations with the appropriate multiplicity factors. By solving Eq. \eqref{running},   the Landau pole can be found straightforwardly: 
\begin{align}
\Lambda_{Landau}=\mu_0 e^{\frac{16 \pi^2}{2 b_R\left( g_R(\mu_0)  \right)^2}}.    
\end{align}
The scale of the Landau pole depends on the value of the coupling $g_R$, at the input scale $\mu_0$. Depending on the choice, both the  $\Lambda_{Landau}<M_{Planck}$ and $\Lambda_{Landau}>M_{Planck}$   scenarios can emerge.

Utilizing the basic set-up defined in Sec. \ref{SEC-02}, we have constructed three different models in Sec. \ref{sec:dirac}, which correspond to three different coefficients $b_R=\{179/3, 56, 731/12\}$ for the Dirac seesaw, simplest one-loop, and Scotogenic models respectively. 
For demonstration purpose, we choose $\mu_0=10$ TeV and show the scale  $\Lambda_{Landau}$ as a function of gauge coupling in Fig. \ref{U1-running} for the three different models discussed in this work. As expected, the higher the value of $g_R$, smaller the  $\Lambda_{Landau}$ gets.

\section{Lepton Flavor Violation}\label{SEC-LFV} 
\begin{figure}[b!]
$$
 \includegraphics[scale=0.56]{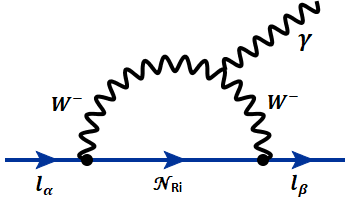} \hspace{0.1 in}
  \includegraphics[scale=0.56]{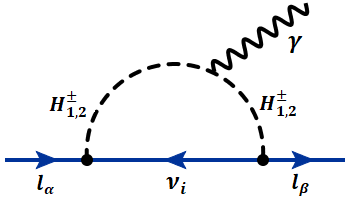} \hspace{0.1 in}
 \includegraphics[scale=0.56]{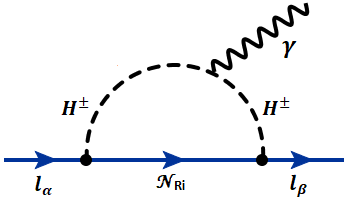} \hspace{0.1 in}
 $$
 \caption{Representative one-loop Feynman diagrams contributing to $\ell_\alpha \to \ell_\beta + \gamma$ processes mediated by charged Bosons in minimal tree-level Dirac seesaw model (left), simplest one-loop Dirac neutrino mass model (middle) and scotogenic Dirac neutrino mass model (right). }
\label{lfv_feyn}
\end{figure}

In this section, we pay special attention to the  charged lepton flavor violation (cLFV)  which is an  integral feature of these Dirac neutrino mass models. These lepton flavor violating processes provide stringent constraints on TeV-scale extensions of the standard model and, as a consequence put restrictions  on the free parameters of our theories.  For the first model we discussed, where neutrino masses are generated via Dirac seesaw mechanism, the cLFV decay rates induced by the neutrino mixings (cf. Fig.  \ref{lfv_feyn}) are highly suppressed by the requirement that the scale of new physics (vector-like fermions $\mathcal{N}_{L,R}$) is at $10^{15}$ GeV to satisfy the neutrino oscillation data, with Yukawa couplings being order one, and hence, are well below the current experimental bounds. Here, we can safely ignore cLFV processes associated with Dirac  seesaw model. On the other hand, in the simplest one-loop Dirac neutrino mass model and in the scotogenic  model, several new contributions appear due to the additional contributions from charged scalars (cf. Fig. \ref{lfv_feyn}), which could lead to sizable cLFV rates. 

 \begin{figure}[htb!]
$$
 \includegraphics[height=6.5cm,width=0.5\textwidth]{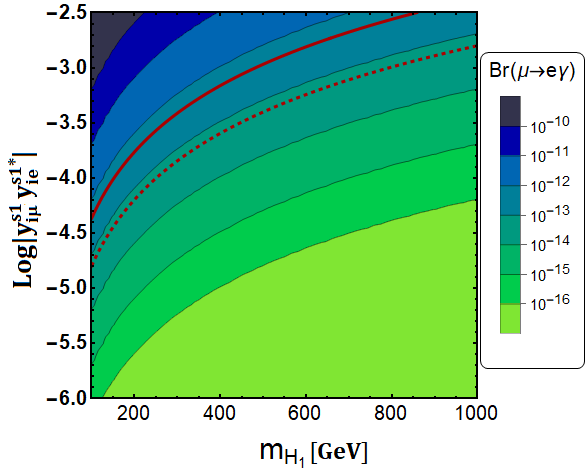} \hspace{0.1 in}
  \includegraphics[height=6.5cm,width=0.5\textwidth]{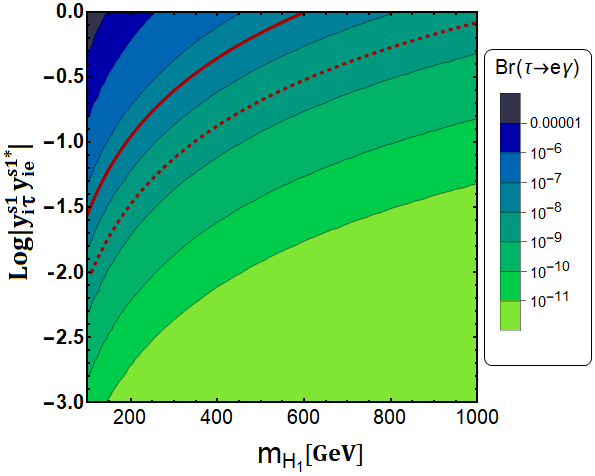} \hspace{0.1 in}
 $$
 $$
 \includegraphics[height=6.5cm,width=0.5\textwidth]{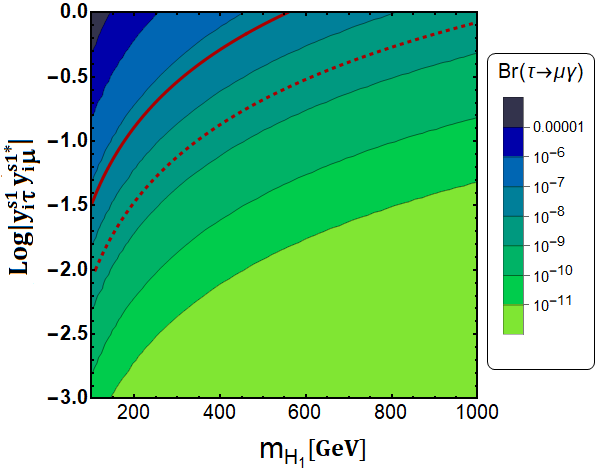} \hspace{0.1 in}
 $$
 \caption{Contour plot for branching ratio predictions for the processes: $\mu \to e + \gamma$ (top left), $\tau \to e + \gamma$ (top right) and  $\tau \to \mu + \gamma$ (bottom) as a function of mass $(m_{H_1})$ and Yukawa plane in simplest one-loop Dirac neutrino mass model. Red solid lines indicate the current bounds on branching ratios and red dashed lines indicate the future projected bounds on  the branching ratios.}
\label{lfv1}
\end{figure}

The cLFV decay processes $\ell_\alpha \to \ell_\beta + \gamma$ arise from one-loop diagrams are shown in Fig.~\ref{lfv_feyn}. 
Let us now focus on the major cLFV processes $\ell_\alpha \to \ell_\beta + \gamma$ in the simplest one-loop Dirac neutrino mass model. Processes of these types are most dominantly mediated by both the $SU(2)_L$ singlet charged scalars ($H_{1,2}^{\pm}$). However, the charged scalar  $S_{1}^{\pm}$ determines the chirality of the initial and final-state charged leptons to be left-handed, whereas $S_{2}^{\pm}$ mediated process fixes the chirality to be right-handed and hence there will be no interference between these two contributions. The Yukawa term $y^{S_1}$ is anti-symmetric in nature, whereas $y^{S_2}$ has completely arbitrary elements in the second and third rows (recall the restriction $y^{S_2}_{i1}=0$). We can always make such a judicious choice that no more than one entry in a given row of $y^{S_2}$ can be large and thus we can suppress the contribution from the charged scalar $H_{2}^{\pm}$ for the cLFV processes. The expression for  $\ell_\alpha \to \ell_\beta + \gamma$  decay rates can be expressed as\footnote{The general expression for this decay rate can be found in Ref.~\cite{Lavoura:2003xp, Babu:2019mfe}. }: 
\begin{equation}
\Gamma\left(\ell_{\alpha} \rightarrow \ell_{\beta}+\gamma\right)=\frac{\alpha}{4\left(16 \pi^{2}\right)^{2}} \frac{m_{\alpha}^{5}}{144}\left[\left(\frac{\cos^2\theta}{m^2_{H_1}}+\frac{\sin^2\theta}{m^2_{H_2}}\right)^2\left|y^{S_1}_{i \alpha} y_{i \beta}^{S_1*}\right|^2 + \left(\frac{\sin^2\theta}{m^2_{H_1}}+\frac{\cos^2\theta}{m^2_{H_2}}\right)^2\left|y^{S_2}_{i \alpha} y_{i \beta}^{S_2*}\right|^2\right].
\end{equation}

\begin{figure}[htb!]
$$
 \includegraphics[height=6.5cm,width=0.5\textwidth]{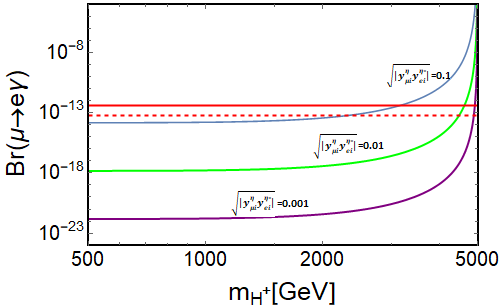} \hspace{0.1 in}
  \includegraphics[height=6.5cm,width=0.5\textwidth]{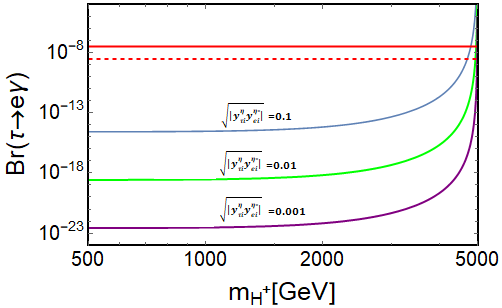} \hspace{0.1 in}
 $$
 $$
 \includegraphics[height=6.5cm,width=0.5\textwidth]{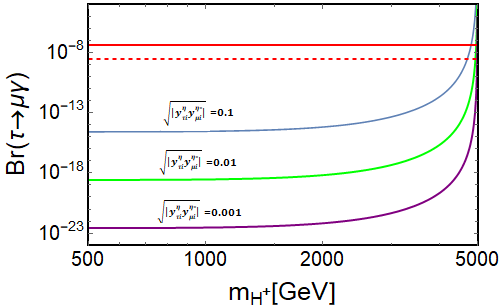} \hspace{0.1 in}
 $$
 \caption{Branching ratio predictions for the processes: $\mu \to e + \gamma$ (top left), $\tau \to e + \gamma$ (top right) and  $\tau \to \mu + \gamma$ (bottom) as a function of mass $(m_{H^+})$   in scotogenic one-loop Dirac neutrino mass model for three benchmark values of Yukawas: $\sqrt{\left|y^{\eta}_{\alpha i} y_{\beta i}^{\eta *}\right| }=10^{-1}, 10^{-2}$ and $10^{-3}$. Red solid lines indicate the current bounds on branching ratios and red dashed lines indicate the future projected bounds on  the branching ratios. }
\label{lfv2}
\end{figure}

In Fig. \ref{lfv1}, we have shown the contour plots for branching ratio predictions for the cLFV processes: $\mu \to e + \gamma$ (top left), $\tau \to e + \gamma$ (top right) and  $\tau \to \mu + \gamma$ (bottom) as a function of mass $(m_{H_1})$ and Yukawa $\left|y^{S_1}_{i \alpha} y_{i \beta}^{S_1*}\right|$ plane in simplest one-loop Dirac neutrino mass model. Red solid lines indicate the current bounds on branching ratios: 4.2 $\times 10^{-13}$ \cite{TheMEG:2016wtm} for the $\mu \to e + \gamma$ (top left) process, 3.3 $\times 10^{-8}$ \cite{Aubert:2009ag} for the $\tau \to e + \gamma$ (top right) process and 4.4 $\times 10^{-8}$ \cite{Aubert:2009ag} for the $\tau \to \mu + \gamma$ (top right) process. Red dashed lines indicate the future projected bounds on  the branching ratios: 6 $\times 10^{-14}$ \cite{fut1} for the $\mu \to e + \gamma$ (top left), 3 $\times 10^{-9}$ \cite{fut2} for the $\tau \to e + \gamma$ (top right)  and 3 $\times 10^{-9}$ \cite{fut2} for the $\tau \to \mu + \gamma$ (top right) processes respectively. For simplicity, we choose $m_{H_2}=m_{H_1}+100$ GeV. As we can see from the Fig. \ref{lfv1}, $\mu \to e + \gamma$ is the most constraining cLFV process in this model. Since this  could lead to sizable rates, it can be tested in the upcoming  experiments.

Similarly, we analyze the major cLFV processes in scotogenic  Dirac neutrino mass model. The representative Feynman diagram for the cLFV process  $\ell_\alpha \to \ell_\beta + \gamma$ is shown in Fig. \ref{lfv_feyn} (right diagram). Here also, charged Higgs $H^{\pm}$, which is the part of the $SU(2)_L$ doublet $\eta$, mainly contributes to the cLFV process $\ell_\alpha \to \ell_\beta + \gamma$ (cf. Fig. \ref{lfv_feyn}). The decay rate for  $\ell_\alpha \to \ell_\beta + \gamma$ solely depends on the two mass terms $m_{H^+}, m_{\mathcal{N}}$ and Yukawa term $y^{\eta}$. The decay width expression for this  process can be written as:
\begin{equation}
\Gamma\left(l_{\alpha} \rightarrow l_{\beta}+\gamma\right)=\frac{\alpha}{4} \frac{\left|y^{\eta}_{\alpha i} y_{\beta i}^{\eta *}\right|^{2}}{\left(16 \pi^{2}\right)^{2}} \frac{\left(m_{\alpha}^{2}-m_{\beta}^{2}\right)^3\left(m_{\alpha}^{2}+m_{\beta}^{2}\right)}{m_{\alpha}^{3}m_{H^{+}}^{4}}\left[f_{B}(t)\right]^{2}.
\end{equation}
Here, $t= m_F^2/m_B^2$, and the function $f_B(t)$ is expressed as \cite{Lavoura:2003xp, Babu:2019mfe}
\begin{equation}
 f_{B}(t) =\frac{2 t^{2}+5 t-1}{12(t-1)^{3}}-\frac{t^{2} \log t}{2(t-1)^{4}}.
\end{equation}

 In Fig. \ref{lfv2}, we have shown the branching ratio predictions for the different cLFV processes: $\mu \to e + \gamma$ (top left), $\tau \to e + \gamma$ (top right) and  $\tau \to \mu + \gamma$ (bottom) as a function of mass $(m_{H^+})$   in scotogenic one-loop Dirac neutrino mass model for three benchmark values of Yukawas: $\sqrt{\left|y^{\eta}_{\alpha i} y_{\beta i}^{\eta *}\right| }=10^{-1}, 10^{-2}$ and $10^{-3}$. For our analysis, we set the vector-like fermion mass $m_{\mathcal{N}}$ to be 5 TeV. The $\mu \to e \gamma $ process imposes  the most stringent bounds. In this set-up, for the Yukawas: $\sqrt{\left|y^{\eta}_{\alpha i} y_{\beta i}^{\eta *}\right| }=10^{-1}, 10^{-2}$ and $10^{-3}$, we get charged Higgs mass bounds to be $m_{H^+}=$  3.1 TeV, 4.6 TeV and 5 TeV respectively. As we can see from Fig. \ref{lfv2}, most of the parameter space in this model is well-consistent with these cLFV processes and which can be testable at the future experiments.
We have shown the future projection reach for these cLFV processes by red dashed lines in Fig. \ref{lfv2}.

\section{Dark Matter Phenomenology }\label{SEC-DM} 
In this section, we briefly discuss the Dark Matter phenomenology in the scotogenic Dirac neutrino mass model. As aforementioned, in this model, a $\mathcal{Z}_2$ subgroup  of the original $U(1)_R$ symmetry remains unbroken that  can stabilize the DM particle. Under this residual symmetry, all the SM particles are even, whereas only the scalars $S,\eta$ and vector-like Dirac fermion $\mathcal{N_{L,R}}$ are odd and  the lightest among these can be the DM candidate. DM phenomenology associated with the neutral component of inert scalar doublet, $\eta$ is extensively studied in Ref. \cite{inert} in a different set-up and corresponding study has been done for the neutral singlet scalar, $S$ in Ref. \cite{Borah:2016zbd, Bhattacharya:2016qsg}. In the following analysis, we consider $\mathcal{N}_1$ to be the lightest among all of these particles, hence  serves as a good candidate for DM (for simplicity we will drop the subscript from $\mathcal{N}_1$ in the following). We aim to study the DM phenomenology associated with the vector-like Dirac fermion $\mathcal{N_{L,R}}$ here. Due to Dirac nature of the dark matter, the phenomenology associated with it is very different from the Majorana fermionic dark matter scenario \cite{Ahriche:2017iar}. 

 \begin{figure}[t!]
$$
 \includegraphics[scale=0.56]{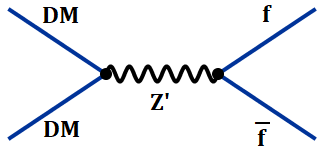} \hspace{0.25 in}
  \includegraphics[scale=0.56]{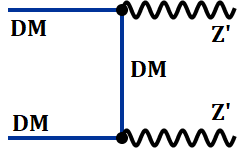} \hspace{0.25 in}
 \includegraphics[scale=0.56]{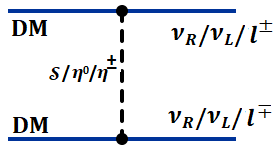} \hspace{0.25 in}
 $$
 \caption{Representative Feynman diagrams for the annihilation of DM particle.}
\label{dm_feyn1}
\end{figure}

 \begin{figure}[b!]
$$
 \includegraphics[height=6.5cm,width=0.5\textwidth]{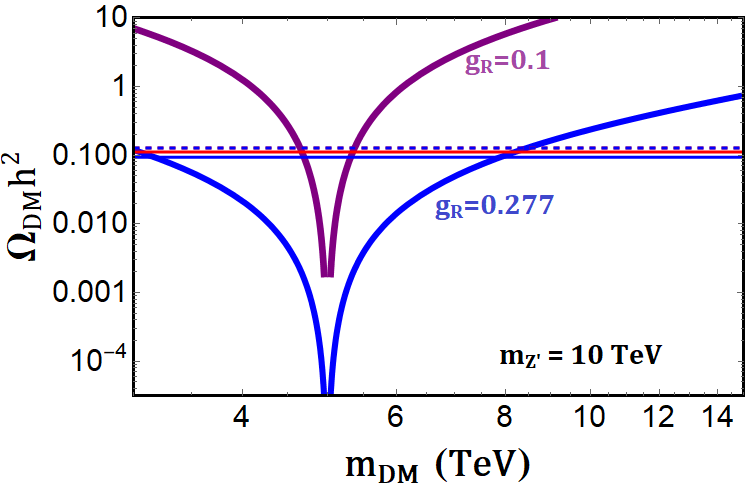} \hspace{0.1 in}
  \includegraphics[height=6.5cm,width=0.5\textwidth]{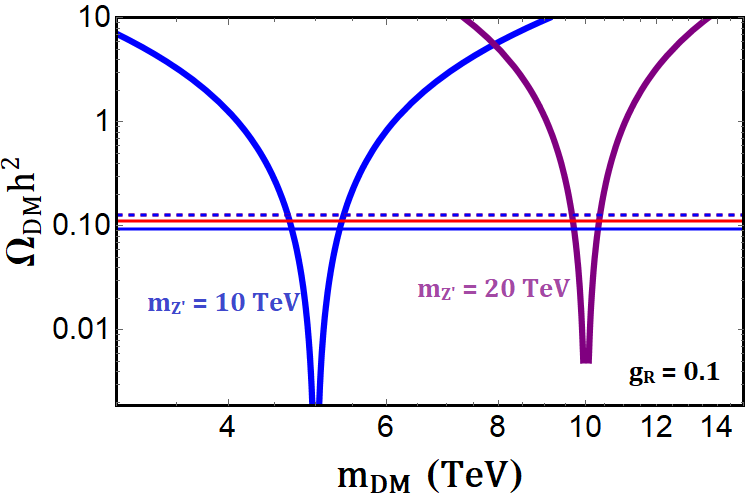} \hspace{0.1 in}
 $$
 \caption{Dark matter relic abundance as a function of dark matter mass $m_{DM}$ for various gauge couplings $g_R$ (left) and $Z'$ boson masses (right). For simplicity, we set $m_{Z'}=10$ TeV (left) and $g_R$= 0.1 (right). Horizontal red and blue lines represent WMAP~\cite{Hinshaw:2012aka} relic density constraint $0.094 \leq \Omega_{\rm DM} h^2 \leq 0.128$  and the PLANCK constraints $0.112 \leq \Omega_{\rm DM} h^2 \leq 0.128$~\cite{Ade:2013zuv} respectively. }
\label{dm1}
\end{figure}

In our case, $\mathcal{N}$  pairs can annihilate through  s-channel $Z'$ exchange process to a pair of SM fermions and right-handed neutrinos. 
Furthermore, if $m_{DM} > m_{Z'}$, then $\mathcal{N}$ may also annihilate directly into pairs of on-shell $Z'$ bosons, which subsequently decay to SM fermions. It can also annihilate to SM fermions and right-handed neutrinos via $t-$ channel scalar ($S, \eta_0,\eta^+$) exchanges. The representative Feynman diagrams for the annihilation of DM particle are shown in Fig. \ref{dm_feyn1}. It is important to mention that for the Majorana fermionic dark matter case, the annihilation rate is $p-$ wave ($\sim v^2$) suppressed since the vector coupling to a self-conjugate particle vanishes, on the contrary, the annihilation rate is not suppressed for the Dirac scenario ($s$-wave).  The non-relativistic form for this annihilation cross-section can be found here \cite{Berlin:2014tja}. In Fig. \ref{dm1}, we analyze the dark matter relic abundance as a function of dark matter mass $m_{DM}$ for various gauge couplings $g_R$ (left) and $Z'$ boson masses (right). Horizontal red and blue lines represent WMAP~\cite{Hinshaw:2012aka} relic density constraint $0.094 \leq \Omega_{\rm DM} h^2 \leq 0.128$  and the PLANCK constraint $0.112 \leq \Omega_{\rm DM} h^2 \leq 0.128$~\cite{Ade:2013zuv} respectively. For simplicity, we set $m_{Z'}=10$ TeV (left)  and provide the relic abundance prediction for two different values of gauge coupling ($g_R$= 0.1 and 0.277). For the right plot in Fig. \ref{dm1}, DM relic abundance is analyzed for two different values of the $Z'$ masses $m_{Z'}=10$ and 20 TeV setting $g_R$= 0.1. As expected, we can satisfy the WMAP~\cite{Hinshaw:2012aka} relic density constraint $0.094 \leq \Omega_{\rm DM} h^2 \leq 0.128$  and the PLANCK constraint $0.112 \leq \Omega_{\rm DM} h^2 \leq 0.128$~\cite{Ade:2013zuv} for most of the parameter space in our model as long as $m_{DM}$ is not too far away from $m_{Z'}/2$ mass. Throughout our DM analysis, we make sure that we are consistent with the SM $Z-$ boson mass correction constraint while choosing specific $g_R$ and $m_{Z'}$ values. 

 \begin{figure}[htb!]
$$
 \includegraphics[scale=0.7]{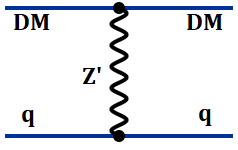} \hspace{0.25 in}
 $$
 \caption{Representative Feynman diagram for the DM-nucleon scattering for   the DM direct detection.}
\label{dm_feyn2}
\end{figure}

 \begin{figure}[htb!]
 \centering
  \includegraphics[scale=0.4]{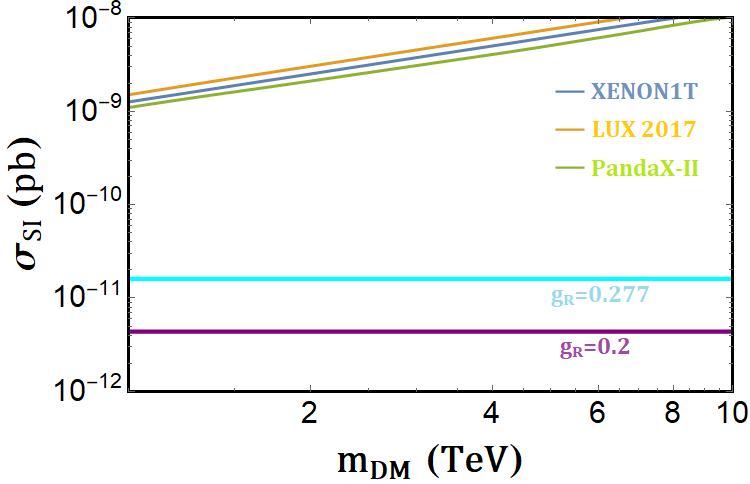}
 \caption{Spin-independent dark matter-nucleon scattering cross-section, $\sigma$ (in pb) as a function of the dark matter mass $m_{DM}$ with different gauge coupling $g_R=0.2, 0.277$. Here we set $m_{Z'}=10$ TeV. Yellow, blue and green color solid lines represent current direct detection cross-section limit from LUX-2017 \cite{Akerib:2016vxi}, XENON1T \cite{Aprile:2017iyp} and PandaX-II (2017) \cite{Cui:2017nnn} experiment respectively. }
\label{dm2}
\end{figure}

In addition to the relic density, we also take into account the constraints from DM direct detection experiments.  In case of Majorana fermionic dark matter, at the tree-level, the spin-independent DM-nucleon scattering cross-section vanishes. However, at the loop-level, the spin-independent operators can be generated and hence it is considerably suppressed. The dominant direct detection signal remains the spin-dependent DM-nucleon scattering cross-section which for the Majorana fermionic dark matter is four times that for the Dirac-fermionic dark matter case. In general, the $Z'$ interactions induce both spin-independent (SI) and spin-dependent (SD) scattering with nuclei. The representative Feynman diagram for the DM-nucleon scattering is shown in Fig. \ref{dm_feyn2}. Particularly, in the scotogenic Dirac neutrino mass model, DM can interact with nucleon through $t-$ channel  $Z'$ exchange. Hence, large coherent spin-independent scattering may occur since both dark matter and the valence quarks of nucleons possess vector interactions with $Z'$  and this process is severely constrained by present direct detection experiment bounds. The DM-nucleon scattering cross-section is estimated in Ref.  \cite{Berlin:2014tja}. In Fig. \ref{dm2}, we analyze the spin-independent dark matter-nucleon scattering cross-section, $\sigma$ (in pb) as a function of the dark matter mass $m_{DM}$ with different gauge coupling $g_R=0.2, 0.277$. For this plot, we set $m_{Z'}=10$ TeV. Yellow, blue and green color solid lines represent current direct detection cross-section limits from LUX-2017 \cite{Akerib:2016vxi}, XENON1T \cite{Aprile:2017iyp} and PandaX-II (2017) \cite{Cui:2017nnn} experiments respectively. As can be seen from Fig. \ref{dm2}, we can satisfy all the present direct detection experiment bounds as long as we are consistent with the other severe bounds on mass $m_{Z'}$ and $g_R$ arising from colliders to be discussed in the next section.

\section{Collider Implications}\label{sec:collider}
Models with extra $U(1)_R$ implies a new $Z^{\prime}$ neutral boson, which contains a plethora of phenomenological implications at colliders. Here we mainly focus on the phenomenology of the heavy gauge boson $Z^{\prime}$ emerging from $U(1)_R$.

\subsection{Constraint on Heavy Gauge Boson $Z^{\prime}$ from LEP}
There are two kinds of $Z^{\prime}$ searches: indirect and direct. In case of indirect searches, one can look for deviations from the SM which might be associated with the existence of a new gauge boson $Z^{\prime}$. This generally involves precision EW measurements below and above the Z-pole. $e^+ e^-$ collision at LEP experiment \cite{lep} above the Z boson mass provides significant constraints on contact interactions involving $e^+ e^-$ and fermion pairs. One can integrate out the new physics and express its influence via higher-dimensional (generally dim-6) operators. For the process $e^+ e^- \to f\bar{f}$, contact interactions can be parameterized by an effective Lagrangian, $\mathcal{L}_{eff}$, which is added to the SM Lagrangian and has the form:

\begin{equation}
{\cal L}_{eff} = \frac{4 \pi}{{\Lambda}^2(1+ \delta_{ef})} \sum_{i,j=L,R} {\eta}_{ij}^{f}  (\bar e_i \gamma^\mu  e_i)(\bar f_j \gamma_\mu  f_j). \label{eq:eff}
\end{equation}
Where $\Lambda$ is the new physics scale, $\delta_{ef}$ is the Kronecker delta function, $f$ indicates all the fermions in the model and $\eta$ takes care of the chirality structure coefficients. The exchange of the new $Z^{\prime}$ boson state emerging from $U(1)_R$ can be stated in a similar way: 
\begin{equation}
{\cal L}_{eff} = \frac{1}{1+ \delta_{ef}} \frac{g_R^2}{M_{Z'}^2} (\bar e \gamma^\mu \mathcal{P}_R e)(\bar f \gamma_\mu \mathcal{P}_R f). \label{eq:eff1}
\end{equation}
Due to the nature of $U(1)_R$ gauge symmetry, the above interaction favors only the right-handed chirality structure. Thus, the constraint on the scale of the contact interaction for the process $e^+ e^- \to l^+l^-$ from LEP measurements \cite{lep} will  
indirectly impose bound on $Z^{\prime}$ mass and the gauge coupling $(g_R)$ that can be translated into:
\begin{equation}
    \frac{M_{Z^{\prime}}}{g_R}\gtrsim 3.59 \rm ~TeV. \label{lepc}
\end{equation}
Other processes such as $e^+ e^- \to c\bar{c}$ and $e^+ e^- \to b\bar{b}$ impose somewhat weaker bounds than the ones quoted in Eq. \ref{lepc}.

\subsection{Heavy Gauge Boson $Z^{\prime}$ at the LHC}
 \begin{figure}[htb!]
$$
 \includegraphics[height=6.5cm,width=0.6\textwidth]{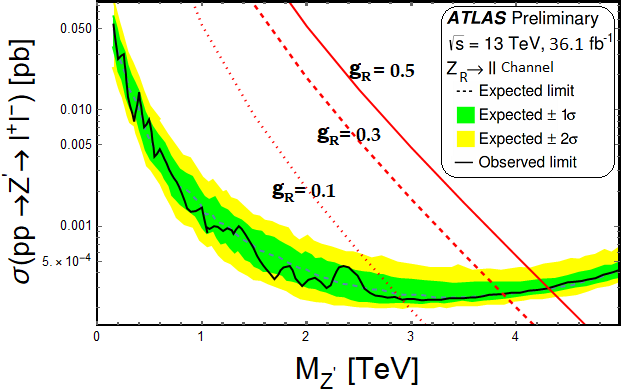} \hspace{0.1 in}
 $$
 \caption{Upper limits at 95$\%$ C.L. on the cross-section for the process $pp \to Z^{\prime} \to l^+ l^- $ as a function of the di-lepton invariant mass using ATLAS results at $\sqrt{s}$ = 13 TeV with 36.1 fb$^{-1}$ integrated luminosity. The black solid line is the observed limit, whereas the green and yellow regions correspond to the 1$\sigma$ and 2$\sigma$ bands on the expected limits. Red  solid (dashed) [dotted] line is for model predicted cross-section for this different values of $U(1)_R$ gauge coupling constant $g_R=0.5 ~(0.3) ~[0.1]$ respectively. }
\label{col1}
\end{figure}
Now we analyze the physics of the heavy neutral gauge boson $Z^{\prime}$ at the Large Hadron Collider (LHC). At the LHC, $Z^{\prime}$ can be resonantly produced via the quark fusion process $q\bar{q} \to Z^{\prime}$ since the coupling of $ Z^{\prime}$ with right-handed quarks ($u_R, d_R$) are not suppressed. After resonantly produced at the LHC, $Z^{\prime}$ will decay into SM fermions and also to the exotic scalars ($S_2^+ S_2^-, \chi \chi $) or fermions ($\mathcal{N}\mathcal{N}$) depending on the model  if kinematically allowed\footnote{Even if we include $Z^{\prime} \to \mathcal{N}\mathcal{N}, \rm ~S_2^+ S_2^- , \rm ~ \chi \chi $ decay modes, the branching fraction ($\sim 4\%$) for $ Z^{\prime} \to e^+ e^- /{\mu}^+ {\mu}^-  $ mode does not change much.}. The present lack of any signal for di-lepton resonances at the LHC dictates the stringent bound on the $Z^{\prime}$ mass and $U(1)_R$ coupling constant $g_R$ in our model as the production cross-section solely depends on these two free parameters. Throughout our analysis, we consider that the mixing $Z-Z^{\prime}$ angle is not very sensitive ($s_X=0$). In order to obtain the constraints on these parameter space,  we use the dedicated search for new resonant high-mass phenomena in di-electron and di-muon final states  using $36.1$ fb$^{-1}$ of proton-proton collision data, collected at $\sqrt{s} = 13$ TeV by the ATLAS collaboration \cite{atlas}. The searches for high mass phenomena in di-jet final states \cite{dijet} will also impose bound on the model parameter space, but it is somewhat weaker than the di-lepton searches due to large QCD background. For our analysis, we implement our models in FeynRules$\char`_$v2.0  package \cite{feynrules} and simulate the events for the process $pp \to Z^{\prime} \to e^+ e^- ({\mu}^+ {\mu}^- ) $ with MadGraph5$\char`_$aMC@NLO$\char`_$v3$\char`_$0$\char`_$1 code \cite{mad}. Then, using parton distribution function (PDF) NNPDF23$\char`_$lo$\char`_$as$\char`_$0130 \cite{nnpdf}, the cross-section and cut efficiencies are estimated. Since no significant deviation from the SM prediction is observed in experimental searches \cite{atlas} for high-mass phenomena in di-lepton final states, the upper limit on the cross-section is derived from the experimental analyses \cite{atlas} using $\sigma \times$  BR $= N_{rec}/(A \times \epsilon \times \int L dt)$, where $N_{rec}$ is the number of reconstructed heavy $Z^{\prime}$ candidate, $\sigma$ is the resonant production cross-section of the heavy $Z^{\prime}$, BR is the branching ratio of $Z^{\prime}$ decaying into di-lepton final states , $A\times \epsilon$ is the acceptance times efficiency of the cuts for the analysis. In Fig. \ref{col1}, we have shown the upper limits on the cross-section at 95$\%$ C.L. for the process $pp \to Z^{\prime} \to l^+ l^- $ as a function of the di-lepton invariant mass using ATLAS results \cite{atlas} at $\sqrt{s}$ = 13 TeV with 36.1 fb$^{-1}$ integrated luminosity. Red  solid, dashed  and dotted lines in Fig. \ref{col1} indicate the model predicted cross-section for three different values of $U(1)_R$ gauge coupling constant $g_R=0.5, ~0.3, ~0.1$ respectively. We find that $Z^{\prime}$ mass should be heavier\footnote{For related works see also \cite{Ekstedt:2016wyi, Bandyopadhyay:2018cwu}.} than $4.4, 3.9 $ and $2.9$ TeV for three different values of $U(1)_R$ gauge coupling constant $g_R=0.5, ~0.3$ and $0.1$.

 \begin{figure}[htb!]
$$
\includegraphics[height=6.5cm,width=0.6\textwidth]{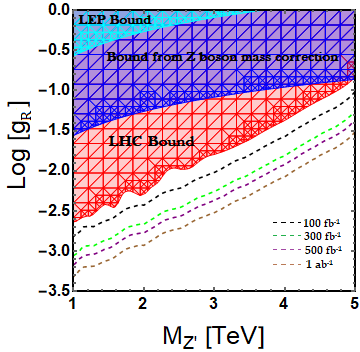} \hspace{0.1 in}
 $$
 \caption{Red meshed zone in $M_{Z^{\prime}} - g_R$ plane  indicates the excluded region from the upper limit on the cross-section for the process $pp \to Z^{\prime} \to l^+ l^- $ at 95$\%$ C.L. using ATLAS results at $\sqrt{s}$ = 13 TeV with 36.1 fb$^{-1}$ integrated luminosity. The cyan meshed zone is excluded from the LEP constraint. The blue meshed zone is excluded from the limit on SM Z boson mass correction: $\frac{1}{3}M_{Z^{\prime}}/g_R > 12.082$ TeV. Black, green, purple and brown dashed lines represent the projected discovery reach at $5 \sigma$ significance  at 13 TeV LHC for 100 fb$^{-1}$, 300 fb$^{-1}$, 500 fb$^{-1}$   and 1 ab$^{-1}$ luminosities. }
\label{col2}
\end{figure}

In Fig. \ref{col2}, we have shown all the current experimental bounds in $M_{Z^{\prime}} - g_R$ plane. Red meshed zone is excluded from the current experimental di-lepton searches \cite{atlas}. The cyan meshed zone is forbidden from the LEP constraint\cite{lep} and the blue meshed zone is excluded from the limit on SM Z boson mass correction: $\frac{1}{3}M_{Z^{\prime}}/g_R > 12.082$ TeV as aforementioned. We can see from Fig.  \ref{col2} that the most stringent bound in $M_{Z^{\prime}} - g_R $ plane is coming from direct $Z^{\prime}$ searches at the LHC. After imposing all the current experimental bounds, we analyze the future discovery prospect of this heavy gauge boson $Z^{\prime}$ within the allowed parameter space in $M_{Z^{\prime}} - g_R $ plane looking at the prompt di-lepton resonance signature at the LHC. We find that a wider region of parameter space in $M_{Z^{\prime}} - g_R$ plane can be tested at the future collider experiment. Black, green, purple and brown dashed lines represent the projected discovery reach at $5 \sigma$ significance  at 13 TeV LHC for 100 fb$^{-1}$, 300 fb$^{-1}$, 500 fb$^{-1}$   and 1 ab$^{-1}$ luminosities. On the top of that, the right-handed chirality structure of $U(1)_R$  can be investigated at the LHC by measuring Forward-Backward (FB)  and top polarization asymmetries in $Z^{\prime} \to t\bar{t}$ mode \cite{fblhc} and which can discriminate our $U(1)_R$ $Z^{\prime}$ interaction from the other $Z^{\prime}$ interactions in $U(1)_{B-L}$ model. The investigation of other exotic decay modes  $( \mathcal{N}\mathcal{N}, \rm ~ \chi \chi, \rm ~ S_2^+ S_2^-  )$ of heavy $Z^{\prime}$ is beyond the scope of this article and shall be presented in a future work since these will lead to remarkable multi-lepton or displaced vertex signature \cite{bog ,jana, jana2, jana3, jana4, jana5, jana6} at the colliders.

\subsection{Heavy Gauge Boson $Z^{\prime}$ at the ILC}
 Due  to  the  point-like  structure  of  leptons  and polarized initial and final state fermions, lepton colliders like ILC will provide much better precision of measurements. The purpose of the $Z^{\prime}$ search at the ILC would be either to help identifying any $Z^{\prime}$ discovered at the LHC or to extend the $Z^{\prime}$ discovery reach (in an indirect fashion) following effective interaction. Even if the mass of the heavy gauge boson $Z^{\prime}$ is too heavy to directly probe at the LHC, we will show that by measuring the process $e^+e^-  \to f^+ f^-$, the effective interaction dictated by Eq.  \ref{eq:eff1} can be tested at the ILC. Furthermore, analysis with the polarized initial states at ILC can shed light on the chirality structure of the effective interaction and thus it can distinguish between the heavy gauge boson $Z^{\prime}$ emerging from $U(1)_R$ extended model  and the $ Z^{\prime}$ from other $U(1)$ extended model such as $U(1)_{B-L}$. The process  $e^+e^-  \to f^+ f^-$ typically exhibits asymmetries in the distributions of the final-state particles isolated by the angular- or polarization-dependence of the differential cross-section. These asymmetries  can thus be utilized as a sensitive measurement of differences in interaction strength and to distinguish a small asymmetric signal at the lepton colliders. In the following, the asymmetries (Forward-Backward asymmetry, Left-Right asymmetry) related to this work will be described in great detail.

\subsubsection{Forward-Backward Asymmetry}
 The differential cross-section in Eq. \ref{dc} is asymmetric in polar angle, leading to a difference of cross-sections for $Z^{\prime}$ decays between the forward and backward hemispheres. 
 Earlier, LEP experiment \cite{lep} used  Forward-backward asymmetries to measure the difference in the interaction strength of the $Z$-boson between left-handed and right-handed fermions, which gives a precision measurement of the weak mixing angle. Here we will show that our framework leads to sizable and distinctive Forward-Backward (FB) asymmetry discriminating from other models and which can be tested at the ILC, since only the right-handed fermions carry non-zero charges under the $U(1)_R$. For earlier analysis of FB asymmetry in the context of other models as well as model-independent analysis  see for example Refs.  \cite{Djouadi:1991sx, DelAguila:1993rw, Cvetic:1995zs, Riemann:1996fk, Leike:1996pj,  Rizzo:1996ce, Babich:1998ri, Leike:1998wr, Casalbuoni:1999mw, Weiglein:2004hn, Godfrey:2005pm, Nomura:2017tih, Nomura:2018mwr}.

 \begin{figure}[htb!]
$$
 \includegraphics[height=6.5cm,width=0.5\textwidth]{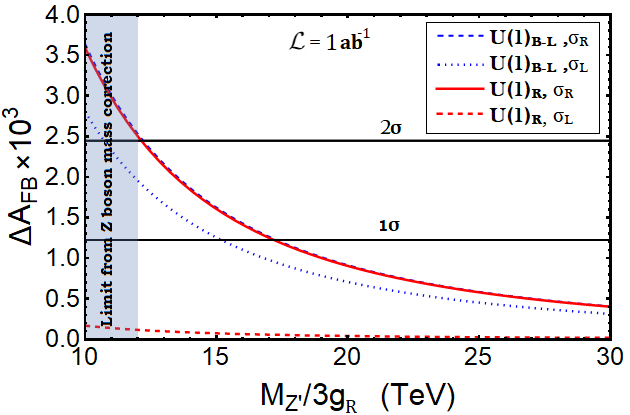} \hspace{0.1 in}
\includegraphics[height=6.5cm,width=0.5\textwidth]{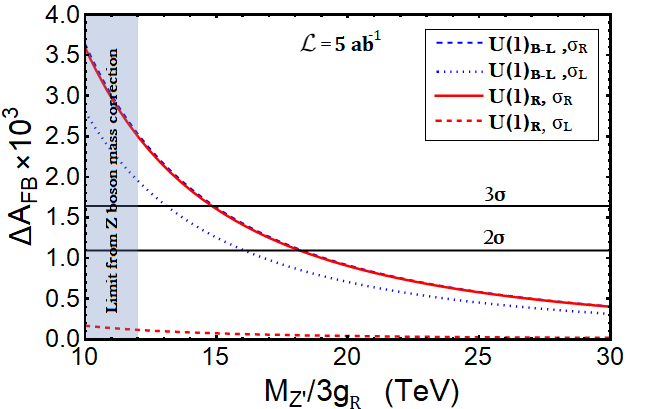} \hspace{0.1 in}
 $$
 \caption{The strength of FB asymmetry $\Delta A_{FB}$ as a function of VEV $v_\chi (= M_{Z^{\prime}}/{3g_R})$ for both left and right-handed polarized cross-sections  of the $e^+ e^- \to \mu^+ \mu^-$ process at the ILC. Red dashed (solid) line represents  $\Delta A_{FB}$ for $U(1)_R$ case for left (right) handed polarized cross-sections  of the $e^+ e^- \to \mu^+ \mu^-$ process, whereas blue dotted (dashed) line indicates  $\Delta A_{FB}$ for $U(1)_{B-L}$ case for left (right) handed polarized cross-sections. Here, we set COM energy of the ILC at $\sqrt{s}$ =  500 GeV with 1 ab$^{-1}$ (left) and 5 ab$^{-1}$ (right) integrated luminosity. Here the horizontal solid black lines correspond to the $1\sigma$ and $2\sigma$ ( $2\sigma$ and $3\sigma$ ) sensitivity for left (right) figure, and the grey shaded region corresponds to excluded region from the SM $Z$ boson mass correction. }
\label{fb1}
\end{figure}

At the ILC, $Z'$ effects have been studied for the following processes:  
\begin{align}
 & e^-(k_1,\sigma_1) + e^+(k_2,\sigma_2) \to e^-   (k_3,\sigma_3) +
 e^+(k_4,\sigma_4) , \\ 
 & e^-(k_1,\sigma_1) + e^+(k_2,\sigma_2) \to \mu^- (k_3,\sigma_3) +
 \mu^+(k_4,\sigma_4) , \\ 
 & e^-(k_1,\sigma_1) + e^+(k_2,\sigma_2) \to \tau^-(k_3,\sigma_3) +
 \tau^+(k_4,\sigma_4), 
\end{align}
where $\sigma_i = \pm1$ are the helicities of initial~(final)-state leptons and $k_i$'s are the momenta. Since  the $e^+ e^- \to \mu^+ \mu^-$ process is the most sensitive one at the ILC, we will focus on this process only for the rest of our analysis. One can write down the corresponding helicity amplitudes as: 
\begin{align}
  & {\cal M}(+-+-) = -e^2\left(1+\cos\theta\right)
 \left[ 1 + c_R^2 \frac{s}{s_Z}
 + \frac{4s}{\alpha (\Lambda_R^e)^2}\right], \\ 
 & {\cal M}(-+-+) = -e^2\left(1+\cos\theta\right)
 \left[ 1 +c_L^2\frac{ s}{s_Z}
 \right], \\ 
 & {\cal M}(+--+) = {\cal M}(-++-) =
 e^2\left(1-\cos\theta\right)\left[1+c_Rc_L\frac{s}{s_Z}\right], \\
 & {\cal M}(++++) = {\cal M}(----) = 0,
\end{align}
where $s=(k_1+k_2)^2=(k_3+k_4)^2$,
 $s_Z=s-m_Z^2+im_Z\Gamma_Z$, 
and $\cos\theta$ indicates the scattering polar angle. $e^2=4\pi\alpha$ with $\alpha=$ QED coupling constant, $c_R=\tan\theta_W$ and $c_L=-\cot2\theta_W$ and $\theta_W$ is the weak
mixing angle. 

For a purely polarized initial state, the differential cross-section is expressed as:
\begin{align}
 \frac{d\sigma_{\sigma_1\sigma_2}}{d\cos\theta} = \frac{1}{32\pi s}
 \sum_{\sigma_3,\sigma_4} \left|{\cal M}_{\{\sigma_i\}}\right|^2.
\end{align}
Then the differential cross-section for the partially polarized initial state with a degree of polarization $P_{e^-}$ for the electron beam and $P_{e^+}$ for the positron beam can be written as  \cite{Djouadi:1991sx, Nomura:2017tih}:
\begin{align}\label{dc}
  \frac{d\sigma(P_{e^-},P_{e^+})}{d\cos\theta} & =
 \frac{1+P_{e^-}}{2} \frac{1+P_{e^+}}{2}\frac{d\sigma_{++}}{d\cos\theta}
 + \frac{1+P_{e^-}}{2}
 \frac{1-P_{e^+}}{2}\frac{d\sigma_{+-}}{d\cos\theta}\nonumber \\
 & + \frac{1-P_{e^-}}{2}
 \frac{1+P_{e^+}}{2}\frac{d\sigma_{-+}}{d\cos\theta} 
 + \frac{1-P_{e^-}}{2}
 \frac{1-P_{e^+}}{2}\frac{d\sigma_{--}}{d\cos\theta}.
\end{align}
One can now define polarized cross-section  $\sigma_{L,R}$ (for the realistic values at the ILC  \cite{Baer:2013cma}) as: 
\begin{align}
 & \frac{d\sigma_R}{d\cos\theta} = 
 \frac{d\sigma(0.8,-0.3)}{d\cos\theta}, \\
 & \frac{d\sigma_L}{d\cos\theta} = 
 \frac{d\sigma(-0.8,0.3)}{d\cos\theta},
\end{align}
Using this one can study the initial state polarization-dependent forward-backward asymmetry as: 
\begin{align}
& A_{FB}\left(\sigma_{L,R}\right) = \frac{N_F\left(\sigma_{L,R}\right) - N_B\left(\sigma_{L,R}\right)}{N_F\left(\sigma_{L,R}\right) + N_B\left(\sigma_{L,R}\right)}, \nonumber \end{align}
where
\begin{align}
& N_{F}\left(\sigma_{L,R}\right) = \epsilon \mathcal{L} \int_{0}^{c_{\rm max}} d \cos \theta \frac{d \sigma\left(\sigma_{L,R}\right)}{d \cos \theta},
\\ 
& N_{B}\left(\sigma_{L,R}\right) = \epsilon \mathcal{L} \int_{-c_{\rm max}}^{0} d \cos \theta \frac{d \sigma\left(\sigma_{L,R}\right)}{d \cos \theta},
\end{align}
where $\mathcal{L}$ represents the integrated luminosity, $\epsilon$ indicates the efficiency of observing the events, and $c_{\rm max}$ is a kinematical cut chosen to maximize the sensitivity. For our analysis we consider $\epsilon = 1$, and $c_{\rm max}=0.95$.
Then we estimate the sensitivity to $Z'$ contribution by:  
\begin{equation}
\Delta A_{FB}\left(\sigma_{L,R}\right) = |A_{FB}^{SM+Z'}\left(\sigma_{L,R}\right)- A_{FB}^{SM}\left(\sigma_{L,R}\right)|,
\end{equation}
where $A_{FB}^{SM+Z'}$ and $A_{FB}^{SM}$ are FB asymmetry originated from both the SM and $Z^{\prime}$ contribution and from the SM case only. Next, it is compared with the statistical error of the asymmetry (in only SM case) $\delta A_{FB}$   \cite{Djouadi:1991sx, Nomura:2017tih}:
\begin{equation}
\delta A_{FB}\left(\sigma_{L,R}\right) = \sqrt{\frac{1-(A_{FB}^{SM}\left(\sigma_{L,R}\right))^2}{N_F^{SM}\left(\sigma_{L,R}\right)+N_B^{SM}\left(\sigma_{L,R}\right)}}.
\end{equation}

In Fig. \ref{fb1}, we analyze the strength of FB asymmetry $\Delta A_{FB}$ as a function of VEV $v_\chi (= M_{Z^{\prime}}/{3g_R})$ for both left and right-handed polarized cross-sections  of the $e^+ e^- \to \mu^+ \mu^-$ process. In order to compare, we have done the analysis for both the cases: $Z^{\prime}$ from both $U(1)_R$ and $U(1)_{B-L}$ cases.  We have considered the center of mass energy for the ILC at $\sqrt{s} = 500 $ GeV and the integrated luminosity $\mathcal{L}$ is set to be 1 ab$^{-1}$ (5 ab$^{-1}$) for the left (right) panel of Fig. \ref{fb1}. The grey shaded region corresponds to excluded region from the SM $Z$ boson mass correction. Red dashed (solid) line represents  $\Delta A_{FB}$ for $U(1)_R$ case for left (right) handed polarized cross-sections  of the $e^+ e^- \to \mu^+ \mu^-$ process, whereas blue dotted (dashed) line indicates  $\Delta A_{FB}$ for $U(1)_{B-L}$ case for left (right) handed polarized cross-sections. From Fig. \ref{fb1}, we find that in case of $U(1)_R$ model, it provides significant difference of $\Delta A_{FB}$   for $\sigma_R$ and $\sigma_L$ due to the right-handed chirality structure of $ Z^{\prime}$ interaction from $U(1)_R$,  while in the case of $U(1)_{B-L}$ model, it provides  small  difference. Hence by comparing the difference of $\Delta A_{FB}$   for differently polarized cross-section $\sigma_R$ and $\sigma_L$ at the ILC, we can easily discriminate the  $ Z^{\prime}$ interaction from $U(1)_R$ and $U(1)_{B-L}$ model. As we can see from Fig. \ref{fb1} that there are significant region for $M_{Z^{\prime}}/3 g_R> 12.082$ TeV which can give more than $2\sigma$ sensitivity for FB asymmetry by looking at $e^+ e^- \to \mu^+ \mu^-$ process at the ILC. We can also expect much  higher sensitivity while combining different final fermionic states such as other leptonic modes ($e^+ e^-, {\tau}^+{\tau}^{−}$) as well as hadronic modes $jj$. Moreover, the sensitivity to $Z^{\prime}$ interactions can be enhanced by analyzing the scattering angular distribution in details, although it is beyond the scope of our paper. 
 
\subsubsection{Left-Right Asymmetry}

 \begin{figure}[htb!]
$$
 \includegraphics[height=6.5cm,width=0.5\textwidth]{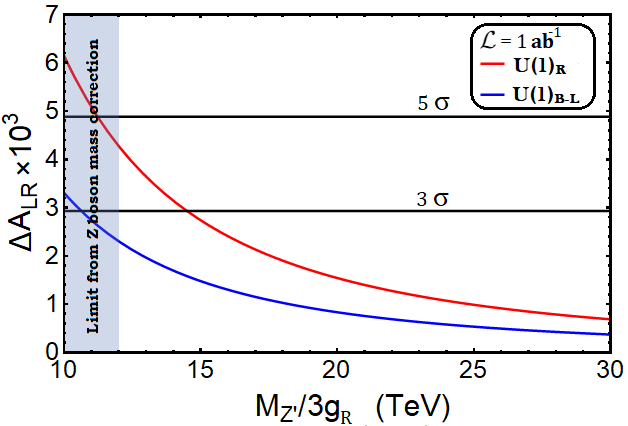} \hspace{0.1 in}
\includegraphics[height=6.5cm,width=0.5\textwidth]{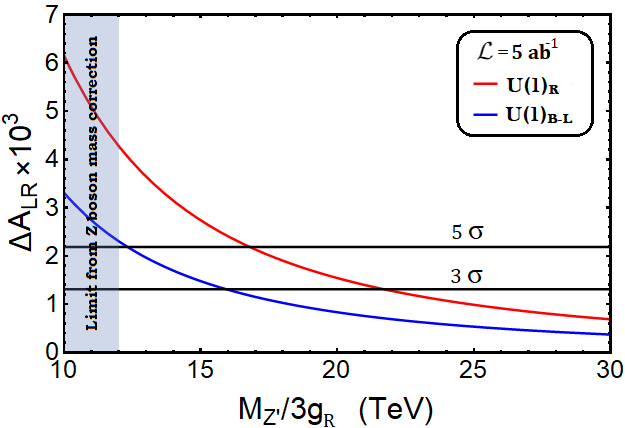} \hspace{0.1 in}
 $$
 \caption{The strength of LR asymmetry $\Delta A_{FB}$ as a function of VEV $v_\chi (= M_{Z^{\prime}}/{3g_R})$ for both left and right-handed polarized cross-sections  of the $e^+ e^- \to \mu^+ \mu^-$ process at the ILC. Red solid line represents  $\Delta A_{LR}$ for $U(1)_R$ case for the $e^+ e^- \to \mu^+ \mu^-$ process, whereas blue solid line indicates  $\Delta A_{LR}$ for $U(1)_{B-L}$ case. Here, we set COM energy of the ILC at $\sqrt{s}$ =  500 GeV with 1 ab$^{-1}$ (left) and 5 ab$^{-1}$ (right) integrated luminosity. Here the horizontal lines corresponding to sensitivity confidence level 3$\sigma$ and 5$\sigma$, and the grey shaded region corresponds to excluded region from the $Z$ boson mass correction. }
\label{lr1}
\end{figure}

The simplest example of the EW asymmetry for an experiment with a polarized electron beam is the left-right asymmetry $A_{LR}$, which measures the asymmetry at the initial vertex.  Since there is no dependence on the final state fermion couplings, one can get an advantage by looking at LR asymmetry at lepton collider.
Another advantage of this LR asymmetry measurement is that it is barely sensitive to the details of the detector. As long as at each value of $\cos{\theta}$, its detection efficiency of fermions is the same as that for anti-fermions, the efficiency effects should be canceled within the ratio because the $Z^{\prime}$ decays into a back-to-back fermion-antifermion pair and about the midplane perpendicular to the beam axis, the detector was designed to be symmetric. For earlier studies  on LR asymmetry in different contexts,  one can see for example Refs.  \cite{Djouadi:1991sx, DelAguila:1993rw, Cvetic:1995zs, Riemann:1996fk, Leike:1996pj,  Rizzo:1996ce, Babich:1998ri, Leike:1998wr, Casalbuoni:1999mw, Weiglein:2004hn, Godfrey:2005pm, Narita:1998rn}. LR asymmetry is defined as:  
\begin{align}
& A_{LR} = \frac{N_L - N_R}{N_L + N_R}, \nonumber 
\end{align}
where $N_L$ is the number of events in which  initial-state particle is left-polarized, while $N_R$ is the corresponding number of right-polarized events.
\begin{align}
& N_{L} = \epsilon \mathcal{L} \int_{-c_{\rm max}}^{c_{\rm max}} d \cos \theta \frac{d \sigma_L}{d \cos \theta},
\\ 
& N_{R} = \epsilon \mathcal{L} \int_{-c_{\rm max}}^{c_{\rm max}} d \cos \theta \frac{d \sigma_R}{d \cos \theta}.
\end{align}
Similarly, one can  estimate the sensitivity to $Z'$ contribution in LR asymmetry by \cite{DelAguila:1993rw, Narita:1998rn, Leike:1996pj}:  
\begin{equation}
\Delta A_{LR} = |A_{LR}^{SM+Z'}- A_{LR}^{SM}|,
\end{equation}
with a statistical error of the asymmetry $\delta A_{{LR}}$ , given \cite{DelAguila:1993rw, Narita:1998rn, Leike:1996pj} as
\begin{equation}
\delta A_{LR} = \sqrt{\frac{1-(A_{LR}^{SM})^2}{N_L^{SM}+N_R^{SM}}}.
\end{equation}

 \begin{figure}[htb!]
$$
 \includegraphics[height=6.5cm,width=0.55\textwidth]{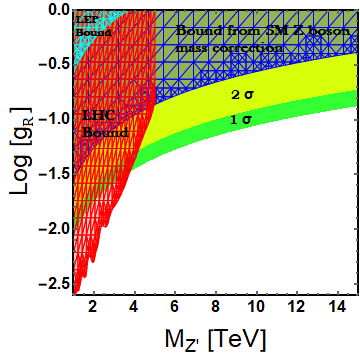} \hspace{0.1 in}
 $$
 \caption{Current existing bounds and projected discovery reach at the ILC in $M_{Z^{\prime}} - g_R$ plane. Green and  yellow shaded zones correspond to sensitivity confidence levels 1$\sigma$ and 2$\sigma$ looking LR asymmetry for $U(1)_R$ extended model at the ILC. Red meshed zone in $M_{Z^{\prime}} - g_R$ plane  indicates the excluded region from the upper limit on the cross-section for the process $pp \to Z^{\prime} \to l^+ l^- $ at 95$\%$ C.L. using ATLAS results at $\sqrt{s}$ = 13 TeV with 36.1 fb$^{-1}$ integrated luminosity. The cyan meshed zone is excluded from the LEP constraint. The blue meshed zone is excluded from the limit on SM Z boson mass correction: $\frac{1}{3}M_{Z^{\prime}}/g_R > 12.082$ TeV. }
\label{fin1}
\end{figure}

In Fig. \ref{lr1}, we analyze the strength of LR asymmetry $\Delta A_{LR}$ for the $e^+ e^- \to \mu^+ \mu^-$ process as a function of VEV $v_\chi (= M_{Z^{\prime}}/{3g_R})$. In order to distinguish $Z^{\prime}$ interaction, we have analysed  both the cases: $Z^{\prime}$ emerging from both $U(1)_R$ and $U(1)_{B-L}$ cases.  We have considered the center of mass energy for the ILC at $\sqrt{s} = 500 $ GeV and the integrated luminosity $\mathcal{L}$ is set to be 1 ab$^{-1}$ (5 ab$^{-1}$) for the left (right) panel of Fig. \ref{lr1}.  The grey shaded region corresponds to excluded region from the SM $Z$ boson mass correction. Red (blue) solid line represents  $\Delta A_{LR}$ for $U(1)_R$ ($U(1)_{B-L}$) case. From Fig. \ref{lr1}, we find that in case of $U(1)_R$ model, it provides remarkably large LR asymmetry $\Delta A_{LR}$   due to the right-handed chirality structure of $ Z^{\prime}$ interaction from $U(1)_R$,  while in case of $U(1)_{B-L}$ model, it gives a smaller contribution. Hence by comparing the difference of $\Delta A_{LR}$  at the ILC, we can easily discriminate the  $ Z^{\prime}$ interaction from $U(1)_R$ and $U(1)_{B-L}$ model. As we can see from Fig. \ref{lr1} that there is a  significant region for $M_{Z^{\prime}}/3 g_R> 12.082$ TeV which can give more than $3\sigma$ sensitivity for LR asymmetry by looking at $e^+ e^- \to \mu^+ \mu^-$ process at the ILC. Even if,  we can achieve $5 \sigma$ sensitivity for a larger parameter space in our framework if integrated luminosity of ILC is upgraded to $5$ ab$^{-1}$. Although, measurement of both the FB and LR asymmetries at the ILC can discriminate $Z^{\prime}$ interaction for $U(1)_R$ model from other $U(1)$ extended models such as $U(1)_{B-L}$ model,  it is needless to mention that the LR asymmetry provides much better sensitivity than the FB asymmetry in our case. In Fig.  \ref{fin1}, we have shown the survived parameter space  in $M_{Z^{\prime}} - g_R$ plane satisfying all existing bounds and which can be probed at the ILC in future by looking at LR asymmetry strength.  Green and  yellow shaded zones correspond to sensitivity confidence levels 1$\sigma$ and 2$\sigma$ by measuring LR asymmetry for $U(1)_R$ extended model at the ILC. For higher  $Z^{\prime}$ mass (above $\sim$ 10 TeV), it is too heavy to directly produce and probe at the LHC looking at prompt di-lepton signature. On the other hand, ILC can probe the heavy  $Z^{\prime}$  effective interaction and LR asymmetry can pin down/distinguish  our $U(1)_R$ model from other existing $U(1)$ extended model for a large region of the parameter space. Thus, $Z^{\prime}$ search at the ILC would help to identify the origin of $Z^{\prime}$ boson as well as  to extend the $Z^{\prime}$ discovery reach following effective interaction.

\section{Constraint from Cosmology} \label{cosmo}
\begin{figure}[t!]
\centering
\includegraphics[scale=0.45]{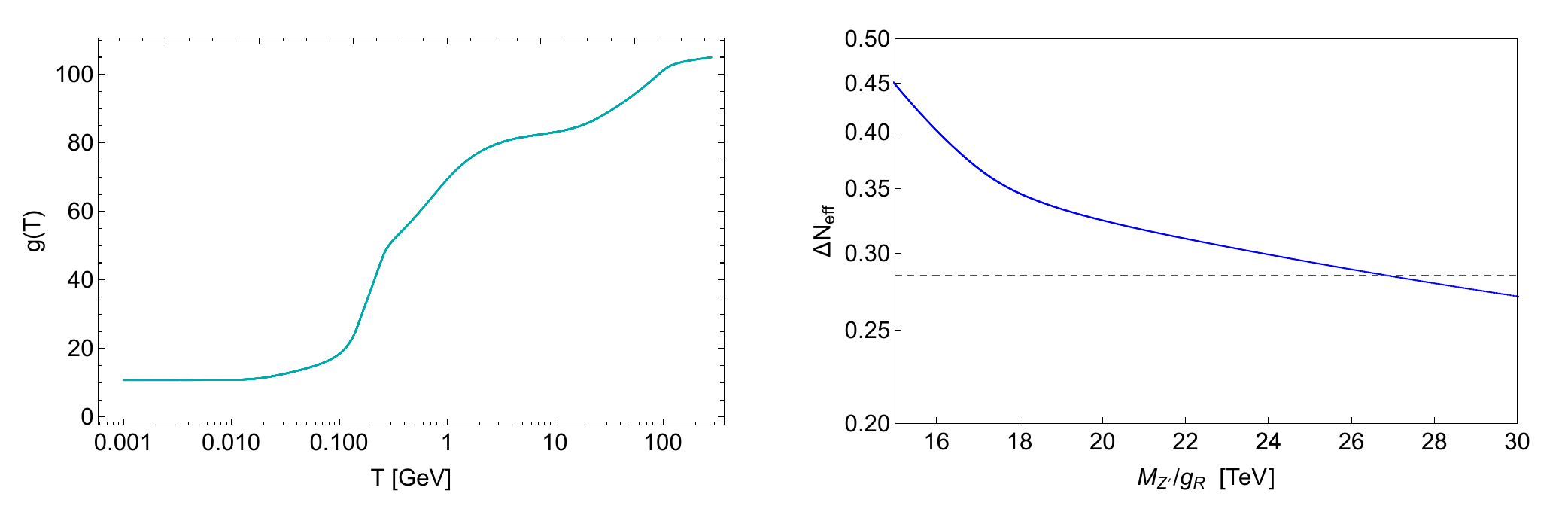} 
\caption{ On the left, we plot the effective number of degrees of freedom as a function of the temperature without including the contribution of the right-handed neutrinos. On the right, we present the contribution of the right-handed neutrinos to $\Delta N_{eff}$ as a function of $M^{\prime}_Z/g_R$. The horizontal dashed red line represents the current upper bound on the shift on the $N_{eff}$ \cite{Aghanim:2018eyx}. 
}\label{gN}
\end{figure}

In the previous section, we have extensively analyzed the collider implications of the new gauge boson  $Z^{\prime}$. In this section, we aim to study the constraints on the mass of the new gauge boson from cosmological measurements and compare with the collider bounds. Since the right-handed neutrinos   carry non-zero $U(1)_R$ charge in our set-up, they couple to the SM sector via the $Z^{\prime}$ boson interactions. Furthermore, since they are either massless or very light,  they contribute to the relativistic degrees of freedom $N_{eff}$, hence in principle can increase the expansion rate of the Universe. Their contribution to this process  is parametrized by $\Delta N_{eff}$ and to compute it we follow the procedure discussed in Ref. \cite{Barger:2003zh}.  After $\nu_R$ states decouple, specifically for $T<T_{dec}^{\nu_L}<T^{\nu_R}_{dec}$ ($T_{dec}^{\nu_{L/R}}$ represents the decoupling temperature of the $\nu_{L/R}$ neutrinos) their total contribution is given by:
\begin{align}
\Delta N_{eff}= N_{\nu_R} \left(  \frac{g(T_{dec}^{\nu_L})}{g(T_{dec}^{\nu_R})} \right)^{4/3},    
\end{align}
here $N_{\nu_R}$ is the number of massless or light right-handed neutrinos, $g(T)$ is the relativistic degrees of freedom at temperature T, with the well-known quantities   $g(T_{dec}^{\nu_L})=43/4$ and $T_{dec}^{\nu_L}=2.3$ MeV  \cite{Enqvist:1991gx}.  For the following computation, we take the temperature-dependent degrees of freedom from the data  listed in Table S2 of Ref. \cite{Borsanyi:2016ksw}, and by utilizing the cubic spline interpolation method, we present $g$ as a function of $T$ in Fig. \ref{gN} (left plot). 

The current cosmological measurement of this quantity is
$N_{eff}=2.99^{+0.34}_{-0.33}$ \cite{Aghanim:2018eyx}, which is completely consistent with the SM prediction $N^{SM}_{eff}=3.045$
\cite{deSalas:2016ztq}. These data limit the contribution of the right-handed neutrinos to be  $\Delta N_{eff}<0.285$.  However, future measurements \cite{Abazajian:2016yjj} can put even tighter  constraints on this deviation $\Delta N_{eff}<0.06$. The right-handed neutrinos decouple from the thermal bath when the  interaction rate drops below the expansion rate of the Universe:
\begin{align}
&\Gamma\left( T_{dec}^{\nu_R}  \right) = H\left( T_{dec}^{\nu_R}  \right).      \label{gn1}  
\end{align}
Here the Hubble expansion parameter is defined as:
\begin{align}
&H^2(T)=T^4 \frac{4 \pi^3}{45 M^2_{Pl}} \left( g(T)+N_{\nu_R}\frac{7}{8}g_{\nu_R}  \right),  \label{gn2} 
\end{align}
where $M_{Pl}$ is the Planck mass and $g_{\nu_R}=2$ is the spin degrees of freedom of the right-handed neutrinos. And the interaction rate that keeps the right-handed neutrinos at the thermal bath is given by:
\begin{align}
&\Gamma(T)=\sum_f \frac{g^2_{\nu_R}}{n_{\nu_R}(T)} \int\frac{d^3p}{(2\pi)^3} \int\frac{d^3q}{(2\pi)^3} f_{\nu_R}(p)   f_{\nu_R}(q) \sigma_f(s) v.   \label{gn3} 
\end{align}
Here, the Fermi-Dirac distribution is $f_{\nu_R}(p)=1/(e^{p/T}+1)$, the number density is $n_{\nu_R}=\left( 3/(2\pi^2) \right)\zeta(3)T^3$, $s=2pq(1-\cos\theta)$ and $v=1-\cos\theta$. Furthermore, the annihilation cross-section $\sigma(\overline{\nu_R}\nu_R\to\overline{f}_if_i)$ is as follows:
\begin{align}
\sigma_f(s)=\sum_f N^f_C Q^2_f \frac{g^4_R}{12\pi\sqrt{s}} \frac{\sqrt{s-4m^2_f}(s+2m^2_f)}{(s-M^2_{Z^{\prime}})^2+\Gamma^2_{Z^{\prime}}M^2_{Z^{\prime}}}.   \label{gn4} 
\end{align}
Where $N^f_C$ and $Q_f$ represent the color degrees of freedom and the charge under the $U(1)_R$ for a fermion $f$ respectively. 

\FloatBarrier

By plugging Eqs. \eqref{gn2}-\eqref{gn4} in Eq. \eqref{gn1} and then  solving numerically, we present our result of $\Delta N_{eff}$ as a function of $M_{Z^{\prime}}/g_R$ in Fig. \ref{gN} (right plot). From this figure, one sees that cosmology provides strong bound on the mass of the new gauge boson based on the associated decoupling temperature of the right-handed neutrinos.  The blue curve corresponds to the contribution of all the three right-handed neutrinos and the red dashed line represents the current experimental upper bound on the deviation of $\Delta N_{eff}$. This bound puts the restriction $M_{Z^{\prime}}/g_R \gtrsim 26.5$ TeV, which is quite stronger than the LEP bound $M_{Z^{\prime}}/g_R \gtrsim 3.59$ TeV, however, lies within the constraint provided by the SM $Z$-boson mass correction $M_{Z^{\prime}}/g_R \gtrsim 36.2$ TeV.   
 The framework presented in this work puts larger bound on the mass of the new gauge boson from cosmology due to large charge assignment of the right-handed neutrinos   compared to the conventional $U(1)_{B-L}$ models with universal charge,  $M_{Z^{\prime}}/g_{B-L} \gtrsim 14$ TeV  \cite{FileviezPerez:2019cyn, Abazajian:2019oqj}.

\section{Conclusions}\label{con}
We believe that the scale of new physics is not  far from the EW scale and a simple extension of the SM should be able to address a few of the unsolved problems of the SM.  Adopting this belief, in this work, we have explored the possibility of one of the most minimal gauge extensions of the SM which is $U(1)_R$ that is responsible for generating Dirac neutrino mass and may also stabilize the DM particle. Cancellations of the gauge anomalies are guaranteed by the presence of the right-handed neutrinos  that   pair up with the left-handed partners to form Dirac neutrinos. Furthermore, this $U(1)_R$ symmetry is sufficient to forbid all the unwanted terms for  constructing  naturally light Dirac neutrino mass models without imposing any additional symmetries by hand. The chiral non-universal structure of our framework  induces asymmetries, such as forward-backward asymmetry  and especially left-right asymmetry that are very distinct compared to  any other $U(1)$ models. By performing detailed phenomenological studies of the associated gauge boson, we have derived the constraints on the $U(1)_R$ model parameter space and analyzed the prospect of its testability at the collider such as at LHC and ILC.  We have shown that a heavy $Z^{\prime}$ (emerging from $U(1)_R$), even if its mass is substantially higher than the center of mass energy available at the  ILC, would manifest itself at tree-level by its propagator effects  producing sizable contributions to the LR asymmetry or FB asymmetry. This can be taken as an initial guide to explore the $U(1)_R$ model at colliders. These models can lead to large lepton flavor violating observables which we have studied and they could give a complementary test for these models. In this work, we have also analyzed the possibility of having a viable Dirac fermionic DM candidate stabilized by the residual discrete symmetry originating  from $U(1)_R$, which connects to SM via $Z'$ portal coupling in a framework that also cater for neutrino mass generation. The DM phenomenology is shown to be crucially dictated by the interaction of $\mathcal{N}$ with $Z'$. Furthermore, we have inspected the constraints coming from the cosmological measurements and compared this result with the different collider bounds. {For a comparison, here we provide a benchmark point by fixing the gauge coupling $g_R=0.056$. With this, the current upper bound on the $Z^{\prime}$ mass is $M_{Z^{\prime}}> 4.25$ TeV from 13 TeV LHC data with $36.1 fb^{-1}$ luminosity, and the future projection reach limit translates into $M_{Z^{\prime}}> 4.67$ TeV with $100 fb^{-1}$ luminosity. Whereas for the same value of the gauge coupling, the ILC has the discovery reach of $4.63$ TeV  at the $2\sigma$ confidence level looking at the left-right asymmetry. The corresponding bounds from LEP, $Z-$boson mass correction and from cosmology are $M_{Z^{\prime}}> 0.2, 2, 1.49$ TeV respectively, which are somewhat weaker compared to LHC and ILC bounds.} To summarize, the presented Dirac neutrino mass models are well motivated and have rich phenomenology. 

\section*{Acknowledgments}
We thank K. S. Babu, Bhupal Dev and S. Nandi for useful discussions. The work of SJ and VPK was in part supported by US Department of Energy Grant Number DE-SC 0016013. The work of SJ was also supported in part by the Neutrino Theory Network Program. SJ thanks the Theoretical Physics Department at Washington University in St. Louis for warm hospitality during the completion of this work.



\FloatBarrier

\end{document}